\documentclass[11pt]{article}

\usepackage[]{inputenc}
\usepackage[T1]{fontenc}
\usepackage{hyperref}

\usepackage{color}

\usepackage{fullpage}

\usepackage{amsmath}
\usepackage{amssymb}
\usepackage[toc,page]{appendix}

\usepackage{amsthm}
\usepackage{float}
\usepackage{wrapfig}
\usepackage{caption}
\usepackage{graphicx}
\usepackage{pgfplots}
\usepackage{subcaption}
\usepackage{tikz,tikz-3dplot}
\tdplotsetmaincoords{80}{45}
\tdplotsetrotatedcoords{-90}{180}{-90}

\tikzset{surface1/.style={draw=blue!70!black, fill=blue!40!white, fill opacity=.6}}
\tikzset{surface2/.style={draw=red!70!black, fill=red!40!white, fill opacity=.6}}
\tikzset{surface3/.style={draw=green!70!black, fill=green!40!white, fill opacity=.6}}

\usetikzlibrary{intersections,fadings,decorations.pathreplacing}

  \pgfplotsset{
        compat=1.8}

\newtheorem*{theorem}{Theorem}

\theoremstyle{definition}
\newtheorem*{definition}{Definition}

\def \be {\begin{equation}}
\def \ee {\end{equation}}
\def \bea {\begin{eqnarray}}
\def \eea {\end{eqnarray}}

\title{Well-posed formulation of Lovelock and Horndeski theories}

\author{Aron D. Kovacs and Harvey S. Reall\\{\small Department of Applied Mathematics and Theoretical Physics, University of Cambridge}\\ {\small Wilberforce Road, Cambridge CB3 0WA, United Kingdom} \\ adk42@cam.ac.uk, hsr1000@cam.ac.uk}

\begin{document}
\maketitle

\begin{abstract}
We study the initial value problem for Lovelock and Horndeski theories of gravity. We show that the equations of motion of these theories can be written in a form that, at weak coupling, is strongly hyperbolic and therefore admits a well-posed initial value problem. This is achieved by introducing a new class of "modified harmonic" gauges for general relativity.
\end{abstract}

\section{Introduction}

Lovelock theories of gravity are the most general theories in which the gravitational field is described by a single metric tensor satisfying a diffeomorphism invariant second order equation of motion \cite{Lovelock1971}. In vacuum, the equation of motion of a Lovelock theory in $d$ spacetime dimensions is:
\be\label{eom_lovelock}
E^{\mu}_{~\nu}\equiv G^\mu{}_\nu +\Lambda \delta^\mu_\nu+\sum\limits_{p\geq 2} k_p~\delta_{\nu \sigma_1\ldots \sigma_{2p}}^{\mu\rho_1\ldots \rho_{2p}}R_{\rho_1\rho_2}{}^{\sigma_1\sigma_2}\ldots R_{\rho_{2p-1}\rho_{2p}}{}^{\sigma_{2p-1}\sigma_{2p}}=0
\ee
where $G_{\mu\nu}$ is the Einstein tensor, $\Lambda$ is the cosmological constant, $k_p$ are dimensionful coupling constants, we have scaled so that the coefficient of the Einstein term is unity\footnote{We will not consider Lovelock theories for which this coefficient vanishes.}, and the generalized Kronecker delta is given by
\be
\delta_{\sigma_1\ldots \sigma_{q}}^{\rho_1\ldots \rho_{q}}=q!~\delta^{\rho_1}_{[\sigma_1}\delta^{\rho_2}_{\sigma_2}\ldots \delta^{\rho_q}_{\sigma_q]}.
\ee
For $d=4$, the antisymmetrization implies that equation \eqref{eom_lovelock} reduces to the vacuum Einstein equation. For $d>4$ Lovelock theories introduce finitely many new terms into the equation of motion. 

The Lovelock theory with $k_p=0$ for $p>2$ is referred to as Einstein-Gauss-Bonnet (EGB) theory. This is of particular interest because it can be motivated by effective field theory (EFT). In EFT, one adds to the Einstein-Hilbert Lagrangian all possible higher derivative scalars built from the metric, with suitable dimensionful coefficients. Consider the EFT for vacuum gravity. There are three independent 4-derivative terms that can appear in the Lagrangian. The first two are $R^2$, $R^{\mu\nu} R_{\mu\nu}$. Since these involve the Ricci tensor, which appears in the equation of motion of the 2-derivative theory, they can be eliminated by field redefinitions \cite{Burgess:2003jk}. The third 4-derivative term can be chosen to be 
\be
\label{LGB}
 {\cal L}_{GB}=\frac14 \delta^{\mu_1\mu_2\mu_3\mu_4}_{\nu_1\nu_2\nu_3\nu_4}R_{\mu_1\mu_2}{}^{\nu_1\nu_2}R_{\mu_3\mu_4}{}^{\nu_3\nu_4}.
\ee
Variation of this term gives the $p=2$ term in the Lovelock equation of motion. Hence, if one truncates the EFT, retaining only the terms with up to $4$ derivatives, then one has EGB theory. Thus, in the absence of matter, EGB theory gives the leading order EFT corrections to General Relativity (GR) in $d>4$ dimensions. 

Horndeski theories are the most general theories of a metric tensor coupled to a scalar field $\phi$, with second order equations of motion, arising from a diffeomorphism-invariant action in $d=4$ spacetime dimensions \cite{Horndeski1974}. A subset of Horndeski theories can be motivated by EFT. We start with a scalar field minimally coupled to gravity and then add to the action all possible higher derivative scalars built from the metric and the scalar field. If we assume a parity symmetry then, by exploiting field redefinitions, the action can be written \cite{Weinberg:2008hq}
\be
\label{4dST}
 S = \frac{1}{16\pi G} \int d^4 x \sqrt{-g} \left( - V(\phi)+ R + X +  \alpha(\phi) X^2 +  \beta(\phi) {\cal L}_{\rm GB}  \right)
\ee
where we have neglected terms with $6$ or more derivatives, $X \equiv -(1/2) g^{\mu\nu} \partial_\mu \phi \partial_\nu \phi$, $V,\alpha,\beta$ are arbitrary functions and ${\cal L}_{GB}$ is given in \eqref{LGB}. We will call this {\it 4-derivative scalar-tensor theory} ($4\partial$ST theory). The special case with $\alpha(\phi)=0$ is called Einstein-dilaton-Gauss-Bonnet (EdGB) theory. $4\partial$ST-theory has second order equations of motion and hence must be a Horndeski theory although some work is required \cite{Kobayashi:2011nu} to rewrite its action in the canonical form of a Horndeski theory.  

A minimal requirement for a Lovelock or Horndeski theory to be viable as a classical field theory is that it should admit a {\it well-posed initial value formulation}. This means that, given initial data satisfying the constraint equations, there exists, for a non-zero time interval, a unique (up to diffeomorphisms) solution of the equations of motion that depends continuously on the initial data. In Appendix \ref{sec:hyperbolicity} we explain why a sufficient condition for well-posedness is that there should exist a formulation of the equations of motion which is {\it strongly hyperbolic} \cite{Kreiss1989,Taylor91,Sarbach2012}. Here, by "formulation", we mean a choice of gauge, and the inclusion of appropriate gauge-fixing terms in the equations of motion. 

Previous work has shown that even {\it weak} hyperbolicity can fail in Lovelock \cite{Reall2014} or Horndeski \cite{Papallo:2017qvl,Ripley:2019hxt,Ripley:2019irj,Ripley:2019aqj} theories once the spacetime curvature and/or scalar field derivatives become large enough that the higher-derivative Lovelock or Horndeski terms in the equation of motion are comparable to the 2-derivative terms. We will refer to this as the {\it strongly coupled} regime of these theories. So in this regime these theories are not viable as classical theories. 

Henceforth we will restrict to the {\it weakly coupled} regime of these theories. By this we mean that the higher derivative terms in the Lovelock or Horndeski equations of motion are small compared to the 2-derivative terms. Note that this is compatible with the fields being strong in the sense of nonlinearities being important e.g. it is compatible with black hole formation provided the black hole is large compared to, say, the length scale defined by the coupling constant $k_2$ in EGB theory. From an EFT perspective, the weakly coupled regime is the only situation in which we would trust EGB or $4\partial$ST theory because once the theory becomes strongly coupled, all of the higher derivative terms that we have neglected would become important. 

The issue of hyperbolicity is gauge-dependent. To obtain a strongly hyperbolic formulation of a gravitational theory one must find a "good gauge" and a good choice of gauge fixing terms. This is in contrast with, say, theories of nonlinear electromagnetism, where one can write the equations of motion in terms of a gauge invariant quantity, namely the field strength (see e.g. \cite{nonlinear_ed}). In General Relativity (GR), the simplest way of establishing well-posedness is to work in {\it harmonic gauge} \cite{Wald:1984rg}, possibly including source terms, i.e., a generalized harmonic gauge. Previous work \cite{Papallo:2017qvl,Papallo:2017ddx} has investigated this class of gauges for Lovelock and Horndeski theories. It was shown that, at weak coupling, these theories are weakly hyperbolic in such gauges. However, for Lovelock theories they are not strongly hyperbolic at weak coupling unless $k_p=0$ for all $p$. Only a small subset of harmonic gauge Horndeski theories are strongly hyperbolic at weak coupling \cite{Papallo:2017qvl,Papallo:2017ddx}. An alternative class of numerical-relativity-inspired gauges was studied in \cite{Kovacs:2019jqj} and used to establish strong hyperbolicity at weak coupling for a larger class of Horndeski theories. Unfortunately it does not appear possible to extend this success to more general Horndeski  theories (such as EdGB) or to Lovelock theories.\footnote{Ref. \cite{Allwright:2018rut} suggests that it might be possible to obtain well-posed equations by adapting a method used for the equations of relativistic viscous hydrodynamics. However, this approach has not yet been applied successfully to gravitational theories.} 

In the absence of a well-posed formulation of the equations of motion, an alternative approach is conventional in EFT. If the coefficients of the higher-derivative terms are proportional to some small parameter $\epsilon$ then one can seek solutions as an expansion in powers of $\epsilon$. For example, this approach has been adopted in recent studies of EdGB theory \cite{Witek:2018dmd,Okounkova:2019zep,Okounkova:2020rqw}. This requires that the solution remains close, globally in time, to a solution of the $\epsilon=0$ theory. However, in practice, small deviations from the $\epsilon=0$ theory may gradually accumulate over time until they become large  (e.g. this could be an orbital phase in a binary black hole spacetime). This would lead to a breakdown of the perturbative approach in a situation where the EFT equations of motion should still be valid\footnote{It has been suggested recently that the so-called dynamical renormalization group method may provide a way around this issue, see e.g. \cite{Okounkova:2017yby} and references therein.}. On the other hand, a well-posed formulation of these equations of motion would be able to accommodate such secular effects \cite{Flanagan:1996gw}.  

In this paper we will introduce a strongly hyperbolic formulation of weakly coupled Lovelock and Horndeski theories. To do this, we will introduce a modification of the usual harmonic gauge condition and gauge-fixing term used in GR. We will prove that our modified harmonic gauge equations of motion for GR are strongly hyperbolic. We will then show that strong hyperbolicity is preserved when we deform the theory by introducing Lovelock or Horndeski terms provided these are small, i.e., provided the theory is weakly coupled. A brief account of this approach, focusing on the theory \eqref{4dST}, appeared in our recent Letter \cite{letter}. The present paper will give a full explanation of our approach for a general Lovelock or Horndeski theory, and a proof of strong hyperbolicity. 

To explain the main idea, consider harmonic gauge GR. In this gauge, the metric satisfies a nonlinear wave equation. Not all solutions of this equation are physical. Unphysical solutions arise from two sources. First there is a residual gauge symmetry, so there are unphysical "pure gauge" solutions. Second, the equation admits solutions which violate the harmonic gauge condition so there are unphysical "gauge-condition violating" solutions. In harmonic gauge GR, both types of unphysical solution propagate at the speed of light, i.e., at the same speed as physical solutions. 

Strong hyperbolicity concerns the behaviour of the theory at high frequency. At high frequency, the above solutions can be regarded (roughly speaking) as waves with different polarizations. Strong hyperbolicity requires that a certain matrix needs to be diagonalizable with real eigenvalues. The eigenvectors of this matrix are essentially the polarization vectors of high frequency waves. Harmonic gauge GR is strongly hyperbolic, so this matrix is diagonalizable. However, its eigenvalues are highly degenerate because the different types of solution all travel with the same speed. In particular, the eigenvalues associated with the "pure gauge" and "gauge-condition violating" polarizations are the same. When we deform the theory by turning on Lovelock or Horndeski terms, the "pure gauge" and "gauge-condition violating" modes continue to propagate at the speed of light and so these eigenvalues remain degenerate. But, generically, if eigenvalues are degenerate then the matrix will not be diagonalizable, and this is why strong hyperbolicity fails in harmonic gauge \cite{Papallo:2017qvl,Papallo:2017ddx}.

Our approach overcomes this problem by introducing a modification of the usual harmonic gauge condition such that the "pure gauge" modes propagate along the null cone of an auxiliary (inverse) metric $\tilde{g}^{\mu\nu}$ instead of the null cone of the physical metric. We implement this gauge condition by adding a gauge-fixing term to the equation of motion for the physical metric. In this gauge-fixing term we introduce another auxiliary (inverse) metric $\hat{g}^{\mu\nu}$. The effect of this is to obtain a new formulation of GR in which the "pure gauge" modes propagate along the null cone of $\tilde{g}^{\mu\nu}$, the "gauge-condition violating" modes propagate along the null cone of $\hat{g}^{\mu\nu}$ and the physical modes propagate along the null cone of $g^{\mu\nu}$. By choosing $\tilde{g}^{\mu\nu}$ and $\hat{g}^{\mu\nu}$ such that these three null cones don't intersect, we ensure that the three different types of mode propagate with different speeds. We will show that this formulation of GR is strongly hyperbolic. Furthermore, the degeneracy discussed above is now absent and so, when we introduce a deformation by turning on Lovelock or Horndeski terms, the theory remains strongly hyperbolic at weak coupling.

This paper is organized as follows. In section \ref{sec:GR} we introduce our modified harmonic gauge formulation of vacuum GR and explain why it admits a well-posed initial value problem. In sections \ref{sec:lovelock} and \ref{sec:horndeski} we extend this formulation to weakly coupled Lovelock and Horndeski theories respectively. Section \ref{sec:discuss} contains further discussion, including the implementation of our formulation in numerical relativity. Appendix \ref{sec:hyperbolicity} explains the connection between strong hyperbolicity and well-posedness. The focus of this paper is on gravitational theories but in Appendix \ref{app:maxwell} we show that our formulation can also be applied to electromagnetism. 

Our conventions agree with those of \cite{Wald:1984rg} unless otherwise stated.

\section{General Relativity in modified harmonic gauge}

\label{sec:GR}

\subsection{The modified harmonic gauge equation of motion}
\label{subsec:GR_eom}

Let $g_{\mu\nu}$ be the physical metric. The vacuum Einstein equation is
\be
 E^{\mu\nu}=0
\ee
where
\be
 E^{\mu\nu} \equiv R^{\mu\nu} - \frac{1}{2} R g^{\mu\nu} + \Lambda g^{\mu\nu} 
\ee
 We introduce an auxiliary (inverse) Lorentzian metric $\tilde{g}^{\mu\nu}$ and define
\be
 H^\mu \equiv \tilde{g}^{\rho\sigma} \nabla_\rho \nabla_\sigma x^\mu = - \tilde{g}^{\rho \sigma} \Gamma^\mu_{\rho \sigma}
\ee
where quantities without tildes are calculated using the metric $g_{\mu\nu}$ (so $H^\mu$ involves both $g_{\mu\nu}$ and $\tilde{g}^{\mu\nu}$). Our modified harmonic gauge condition is
\be
\label{gauge_cond}
H^\mu=0
\ee
This is a linear wave equation for $x^\mu$, which admits a well-posed initial value problem for initial data prescribed on a surface $\Sigma$ that is spacelike w.r.t. $\tilde{g}^{\mu\nu}$. So, at least locally, coordinates can be chosen to satisfy this gauge condition, just as for conventional harmonic gauge \cite{Wald:1984rg}.

Now introduce another auxiliary (inverse) Lorentzian metric $\hat{g}^{\mu\nu}$ and define
\be\label{gauge_fixing}
 {E}^{\mu\nu}_{\rm mhg} = E^{\mu\nu} + \hat{P}_\alpha{}^{\beta \mu \nu} \partial_\beta H^\alpha 
\ee
where
\be
\label{hatP}
 \hat{P}_\alpha{}^{\beta \mu \nu} = \delta_\alpha^{(\mu} \hat{g}^{\nu)\beta}- \frac{1}{2} \delta_\alpha^\beta \hat{g}^{\mu\nu} 
\ee
Our modified harmonic gauge equation of motion is
\be
\label{def_har}
 {E}^{\mu\nu}_{\rm mhg} = 0
\ee
We have three inverse metrics $g^{\mu\nu}$, $\tilde{g}^{\mu\nu}$ and $\hat{g}^{\mu\nu}$. The inverse of $g^{\mu\nu}$ is denoted, as usual, by $g_{\mu\nu}$ and index raising and lowering is always performed with $g$. When we need to refer to the inverse of $\hat{g}^{\mu\nu}$ (say) we will write $(\hat{g}^{-1})_{\mu\nu}$. The usual harmonic gauge formulation of GR is obtained by choosing $\hat{g}^{\mu\nu} = \tilde{g}^{\mu\nu} = g^{\mu\nu}$.

\begin{figure}[H]
\centering
\begin{subfigure}{0.5\linewidth}
\centering
 \includegraphics[scale=1.0]{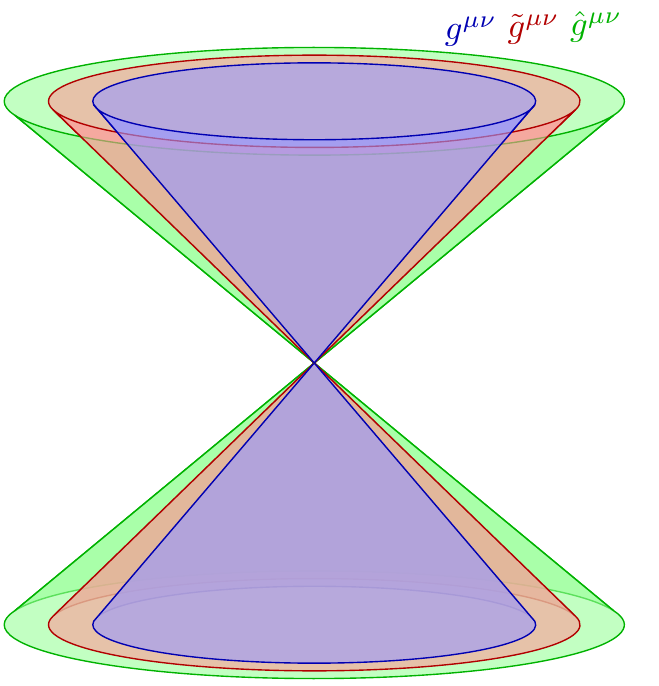}
\caption{}
\label{fig:cones1}
\end{subfigure}%
\begin{subfigure}{0.5\linewidth}
\centering
 \includegraphics[scale=1.0]{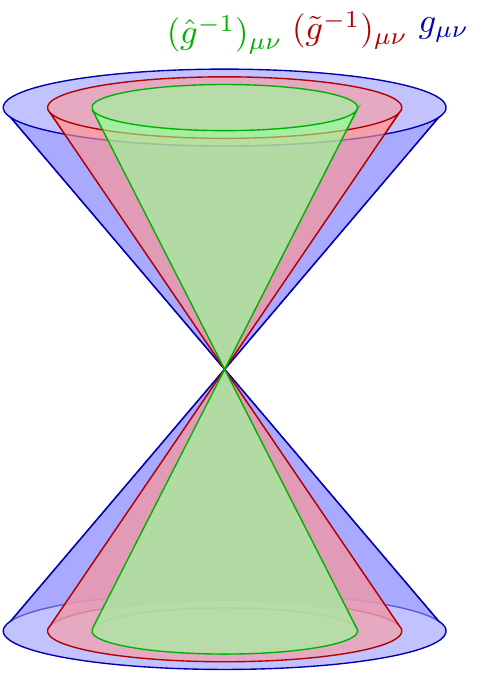}
\caption{}
\label{fig:cones2}
\end{subfigure}
\caption{(a) Cotangent space at a point, showing the
null cones of $g^{\mu\nu}$, $\tilde{g}^{\mu\nu}$ and $\hat{g}^{\mu\nu}$. 
(b) Tangent space at a point, showing the null cones of $g_{\mu\nu}$, $(\tilde{g}^{-1})_{\mu\nu}$ and $(\hat{g}^{-1})_{\mu\nu}$.}
\label{fig:cones}
\end{figure}

We will assume that $\hat{g}^{\mu\nu}$ is chosen so that the causal cone of $g^{\mu\nu}$ (in the cotangent space) lies strictly inside the causal cone of $\hat{g}^{\mu\nu}$, so that any covector that is causal w.r.t. $g^{\mu\nu}$ is timelike w.r.t. $\hat{g}^{\mu\nu}$. See Fig. \ref{fig:cones1}. This implies that the causal cone of $(\hat{g}^{-1})_{\mu\nu}$ (in the tangent space) lies strictly inside the causal cone of $g_{\mu\nu}$ (Fig. \ref{fig:cones2}) so any smooth curve that is causal w.r.t. $(\hat{g}^{-1})_{\mu\nu}$ is timelike w.r.t. $g_{\mu\nu}$. This implies that any point in the domain of dependence $D(\Sigma)$ of a partial Cauchy surface $\Sigma$ w.r.t. $g_{\mu\nu}$ is also in the domain of dependence $\hat{D}(\Sigma)$ of $\Sigma$ w.r.t. $(\hat{g}^{-1})_{\mu\nu}$. In other words, $D(\Sigma) \subset \hat{D}(\Sigma)$. 

We will also assume that $\tilde{g}^{\mu\nu}$ is chosen so that the causal cones of the three inverse metrics form a nested set as in Fig. \ref{fig:cones1}, with the null cones of $\hat{g}^{\mu\nu}$ and $\tilde{g}^{\mu\nu}$ lying outside the null cone of $g^{\mu\nu}$. This implies that a surface $\Sigma$ that is spacelike w.r.t. $g^{\mu\nu}$ is also spacelike w.r.t. $\hat{g}^{\mu\nu}$ and $\tilde{g}^{\mu\nu}$.

In Fig \ref{fig:cones} we have drawn the null cone of $\tilde{g}^{\mu\nu}$ inside that of $\hat{g}^{\mu\nu}$ but we could also choose it to lie outside. What is important is that these null cones do not intersect and that they both lie outside that of $g^{\mu\nu}$.\footnote{\label{fn:nr} In section \ref{sec:discuss} we will comment on how the latter assumption might be relaxed in numerical relativity applications.}

Since the causal cones of the three metrics form a nested set, there are no subtleties with defining time orientations for the unphysical auxiliary metrics. Given a time orientation for the physical metric $g_{\mu\nu}$ we define the future (past) causal cone of $(\hat{g}^{-1})_{\mu\nu}$ to be the one inside the future (past) causal cone of $g_{\mu\nu}$ and similarly for $(\tilde{g}^{-1})_{\mu\nu}$.

In Appendix \ref{app:maxwell} we explain how our modified harmonic gauge condition and gauge-fixing procedure can also be applied to Maxwell theory, which gives a "modified Lorenz gauge" formulation of Maxwell's equations.

\subsection{Propagation of the gauge condition}

\label{subsec:propagation}

Our first task is to show that solutions of \eqref{def_har} are also solutions of the vacuum Einstein equation provided that the initial data satisfies the constraint equations and the modified harmonic gauge condition. The argument follows closely the usual argument for harmonic gauge GR \cite{Wald:1984rg}. Given a solution $g_{\mu\nu}$ of \eqref{def_har} on a manifold $M$, the contracted Bianchi identity gives 
\be
\label{Heq}
 0 = \nabla_\nu {E}^{\mu\nu}_{\rm mhg} = \frac{1}{2} \hat{g}^{\alpha \beta} \partial_\alpha \partial_\beta H^\mu + \ldots
\ee
where the ellipsis denotes terms linear in first derivatives of $H^\rho$. Thus the modified harmonic gauge equation of motion implies that $H^\mu$ satisfies a linear wave equation with principal symbol $(1/2) \hat{g}^{\alpha \beta} \xi_\alpha \xi_\beta$. Let $\Sigma \subset M$ be a surface that is spacelike with future-directed unit normal $n^\mu$ w.r.t. $g^{\mu\nu}$. Then $\Sigma$ is also spacelike w.r.t. $\hat{g}^{\mu\nu}$ so \eqref{Heq} admits a well-posed initial value problem for initial data $H^\mu$ and $\hat{g}^{\nu \rho} n_\nu \partial_\rho H^\mu$ prescribed on $\Sigma$. If $H^\mu$ and $\hat{g}^{\nu \rho} n_\nu \partial_\rho H^\mu$ vanish on $\Sigma$ then it follows from well-posedness of the initial value problem for \eqref{Heq} that $H^\mu$ vanishes throughout the domain of dependence $\hat{D}(\Sigma) \subset M$. Hence $(M,g)$ will satisfy the vacuum Einstein equation $E^{\mu\nu}=0$ in $\hat{D}(\Sigma)$. Since $D(\Sigma) \subset \hat{D}(\Sigma)$, it then follows that $(M,g)$ satisfies the Einstein equation in $D(\Sigma)$. 

Now consider the initial value problem for \eqref{def_har}. In GR, initial data is a triple $(\Sigma,h_{ij},K_{ij})$ where $\Sigma$ is a 3-manifold and, in some chart $x^i$ on $\Sigma$, $h_{ij}$ and $K_{ij}$ are the components of a Riemannian metric and a symmetric tensor on $\Sigma$. These must satisfy the usual constraint equations of GR. We now parameterize the metric $g_{\mu\nu}$ in terms of a lapse function and shift vector in the usual way, which ensures that surfaces of constant $x^0$ are spacelike w.r.t. $g_{\mu\nu}$ and hence also w.r.t. $\hat{g}^{\mu\nu}$ and $\tilde{g}^{\mu\nu}$. At $x^0=0$ the lapse and shift can be chosen arbitrarily. Given a choice of lapse and shift, the values of $g_{ij}$ and $\partial_0 g_{ij}$ at $x^0=0$ are fixed by requiring that the surface $x^0=0$ matches the data on $\Sigma$, i.e., it has induced metric $h_{ij}$ and extrinsic curvature $K_{ij}$. 

The time derivatives of the lapse and shift at $x^0=0$ are fixed by requiring that $H^\mu=0$ at $x^0=0$. This is possible because the equation $H_i=0$ has the form $\tilde{g}^{00} \partial_0 g_{0i} = \ldots$ where the ellipsis denotes terms not involving $\partial_0 g_{0\mu}$. The surface $x^0=0$ is spacelike w.r.t. $\tilde{g}^{\mu\nu}$ so $\tilde{g}^{00} \ne 0$ hence $\partial_0 g_{0i}$ can be chosen to ensure that $H_i=0$. The equation $H_0=0$ then has the form $\tilde{g}^{00} \partial_0 g_{00} = \ldots$ where the ellipsis is independent of $\partial_0 g_{00}$. Hence $\partial_0 g_{00}$ can be chosen to ensure that $H_0=0$. 

We have specified initial data $(g_{\mu\nu},\partial_0 g_{\mu\nu})$ at $x^0=0$ that matches the initial data on $\Sigma$ and satisfies $H^\mu=0$ at $x^0=0$. We can now identify $\Sigma$ with the surface $x^0=0$.

The initial data satisfies the constraint equations of GR so $E^{\mu 0}=0$ at $x^0=0$. Evaluating the $0\mu$ components of \eqref{def_har} at $x^0=0$ and using the vanishing of the tangential derivative $\partial_i H^\mu$ at $x^0=0$, we obtain $\partial_0 H^\mu=0$ at $x^0=0$. Hence all first derivatives of $H^\mu$ vanish at $x^0=0$ so $\hat{g}^{\nu \rho} n_\nu \partial_\rho H^\mu=0$ on $\Sigma$. 

In summary, we have shown that we can choose the initial time derivative of the lapse and shift such that $H^\mu =\hat{g}^{\nu \rho} n_\nu \partial_\rho H^\mu=0$ on $\Sigma$. Hence, if $(M,g)$ is a solution of \eqref{def_har} that matches our initial data $(g_{\mu\nu},\partial_0 g_{\mu\nu})$ on $\Sigma$, then $g_{\mu\nu}$ will then satisfy the vacuum Einstein equation throughout $D(\Sigma) \subset M$.

For technical reasons we will explain later, we will demand that the initial lapse and shift are chosen so that $\partial/\partial x^0$ is timelike w.r.t. all three metrics (although this condition may not be necessary for well-posedness). If this condition is satisfied initially then, by continuity, it will hold in a neighbourhood of the initial surface. 

\subsection{Strong hyperbolicity}\label{sec:GR_strong_hyperbolicity}

In this section we will show that the modified harmonic gauge equation of motion \eqref{def_har} admits a well-posed initial value problem. In Appendix \ref{sec:hyperbolicity} we review how strong hyperbolicity of an equation guarantees well-posedness of the initial value problem. So in this section we will establish well-posedness by showing that \eqref{def_har} is strongly hyperbolic. Strong hyperbolicity is a property of the 2nd derivative terms in the equation of motion, i.e., of the principal symbol.\footnote{We assume that the reader is familiar with the definition of the principal symbol and characteristic covectors. See e.g. \cite{Sarbach2012} for an introduction to these ideas.}

Let $\xi_\mu$ be an arbitrary covector. The principal symbol of \eqref{def_har}, acting on a symmetric tensor $t_{\mu\nu}$, is defined by the replacement $\partial_\mu \partial_\nu g_{\rho \sigma} \rightarrow \xi_\mu \xi_\nu t_{\rho \sigma}$ in the terms involving 2nd derivatives. The result is
\be
\label{principal}
 {\cal P}(\xi)^{\mu\nu\rho\sigma} t_{\rho \sigma} =  {\cal P}_\star(\xi)^{\mu\nu\rho\sigma} t_{\rho \sigma} + {\cal P}_{\rm GF}(\xi)^{\mu\nu\rho\sigma} t_{\rho \sigma} 
 \ee
where we have decomposed the RHS into a part arising from the Einstein tensor and a part arising from the gauge-fixing term in \eqref{def_har}. The part arising from the Einstein tensor is
\be
\label{princ_star}
 {\cal P}_\star(\xi)^{\mu\nu\rho\sigma} t_{\rho \sigma} =
-\frac{1}{2} g^{\gamma\delta} \xi_\gamma \xi_\delta P^{\mu\nu\rho\sigma} t_{\rho \sigma} + P_\alpha{}^{\gamma \mu\nu} \xi_\gamma g^{\alpha \beta} P_\beta{}^{ \delta \rho \sigma} \xi_\delta t_{\rho \sigma} 
\ee

\noindent
and the part arising from the gauge-fixing term is
\be
\label{princ_GF}
  {\cal P}_{\rm GF}(\xi)^{\mu\nu\rho\sigma} t_{\rho \sigma} =- \hat{P}_\alpha{}^{\gamma \mu\nu} \xi_\gamma g^{\alpha \beta} \tilde{P}_\beta{}^{\delta \rho \sigma} \xi_\delta t_{\rho \sigma}
\ee
where we have defined, in analogy with \eqref{hatP},
\be
  P_\alpha{}^{\beta \mu \nu} = \delta_\alpha^{(\mu} g^{\nu)\beta}- \frac{1}{2} \delta_\alpha^\beta g^{\mu\nu}  \qquad  \tilde{P}_\alpha{}^{\beta \mu \nu} = \delta_\alpha^{(\mu} \tilde{g}^{\nu)\beta}- \frac{1}{2} \delta_\alpha^\beta \tilde{g}^{\mu\nu} 
\ee
In conventional harmonic gauge ($\hat{g}^{\mu\nu} = \tilde{g}^{\mu\nu} = g^{\mu\nu}$), the gauge fixing term cancels the second term of \eqref{princ_star} but this is no longer the case in our modified harmonic gauge. 

We will use indices $I,J, \ldots$ to refer to a basis for symmetric tensors, i.e., we will sometimes write $t_I$ instead of $t_{\mu\nu}$. Such indices take values from $1$ to $N = d(d+1)/2$, where $d$ is the spacetime dimension. We can then view ${\cal P}(\xi)$ as a $N \times N$ matrix $P(\xi)^{IJ}$. 
If we do this then the matrix ${\cal P}_\star(\xi)^{IJ}$ is {\it symmetric}. Since ${\cal P}(\xi)$ is quadratic in $\xi_\mu$ we have
\be\label{P_in_IJ}
 {\cal P}(\xi)^{IJ} = {\cal P}^{IJ \gamma \delta} \xi_\gamma \xi_\delta
\ee
where $ {\cal P}^{IJ \gamma \delta}$ can be read off from the above expressions. In coordinates $x^\mu= (x^0,x^i)$ we can write
\be\label{def_ABC_1}
 {\cal P}(\xi)^{IJ} = \xi_0^2 A^{IJ} + \xi_0 B^{IJ} + C^{IJ}
\ee
where
\be\label{def_ABC_2}
 A^{IJ} = {\cal P}^{IJ00} \qquad B^{IJ} = 2\xi_i {\cal P}^{IJ0i} \qquad C^{IJ} = \xi_i \xi_j {\cal P}^{IJij}
\ee
Note that $\xi_i$ are the components of the pull-back of $\xi_\mu$ to the surfaces of constant $x^0$. 

We can write $A^{IJ} = A^{IJ}_\star + A^{IJ}_{\rm GF}$ etc, and the quantities with a star subscript are symmetric matrices. As explained in section \ref{subsec:propagation}, we can arrange that surfaces of constant $x^0$ are spacelike w.r.t. $g^{\mu\nu}$, at least in a neighbourhood of our initial value surface. This implies that these surfaces are also spacelike w.r.t. $\hat{g}^{\mu\nu}$ and $\tilde{g}^{\mu\nu}$. We will show below that a covector is characteristic if, and only if, it is null w.r.t. one of these three inverse metrics. Since $dx^0$ is timelike w.r.t. $g^{\mu\nu}$, it is timelike w.r.t. all three inverse metrics. It follows that $dx^0$ is non-characteristic, which implies that surfaces of constant $x^0$ are non-characteristic and hence the matrix $A^{IJ}$ is invertible.

As reviewed in Appendix \ref{sec:hyperbolicity}, to define strong hyperbolicity we introduce a $(2N) \times (2N)$ real matrix depending on the (real) spatial components $\xi_i$ of $\xi_\mu$ (as well as the spacetime coordinates $x^\mu$ but we suppress this dependence):
\be
\label{Mdef}
 M(\xi_i) = \left( \begin{array}{cc} 0 & I \\ -A^{-1} C(\xi_i) & -A^{-1} B(\xi_i) \end{array} \right)  
\ee
We assume there exists a smooth Riemannian inverse metric $G^{ij}$ on surfaces of constant $x^0$. For example, our condition that $\partial/\partial x^0$ is timelike implies that $g^{ij}$ is positive definite so we could choose $G^{ij} = g^{ij}$. We say that $\xi_i$ is a unit covector if $G^{ij} \xi_i \xi_j = 1$. 

Strong hyperbolicity is the statement that, for any (real) unit covector $\xi_i$, the matrix $M(\xi_i)$ admits a {\it symmetrizer}: a positive definite Hermitian matrix $K(\xi_i)$ such that
\be
\label{sym_def}
  K(\xi_i) M(\xi_i) = M(\xi_i)^\dagger K(\xi_i) 
\ee 
The symmetrizer must satisfy the condition that it depends {\it smoothly} on $\xi_i$ and also on the spacetime coordinates $x^\mu$ that we have suppressed above.

Strong hyperbolicity implies that $M(\xi_i)$ is diagonalizable with real eigenvalues. Conversely, if $M(\xi_i)$ is diagonalizable with real eigenvalues then one can construct a symmetrizer provided the eigenvectors of $M(\xi_i)$ depend smoothly on the unit vector $\xi_i$. The symmetrizer is $(S^{-1})^\dagger S^{-1}$ where $S$ is the matrix whose columns are the eigenvectors. 

In appendix \ref{sec:hyperbolicity} we explain why strong hyperbolicity guarantees well-posedness of the initial value problem. The argument presented there assumes $\det M \ne 0$, which is equivalent to $\det C \ne 0$, which is equivalent to the condition that a covector of the form $(0,\xi_i)$ is never characteristic. This is guaranteed by our condition that $\partial/\partial x^0$ is timelike w.r.t. all three metrics, since this implies $\xi_0 \ne 0$ for any covector $\xi_\mu$ that is null w.r.t. one of the three metrics, as we will show is the case for a characteristic covector.

We will now determine the eigenvalues and eigenvectors of $M(\xi_i)$. 
If $\xi_0$ is an eigenvalue of $M(\xi_i)$ then the eigenvalue equation reduces to the condition that the eigenvector is of the form $(t_I, \xi_0 t_I)^T$ where
\be
\label{characteristic}
 {\cal P}(\xi)^{\mu\nu\rho\sigma} t_{\rho\sigma} = 0
\ee
with $\xi_\mu = (\xi_0,\xi_i)$. This equation states that $\xi_\mu$ is characteristic, with $\xi_0$ a root of the {\it characteristic polynomial} $\det {\cal P}(\xi)$. This is a polynomial of degree $2N$ in $\xi_0$ hence there are $2N$ (possibly degenerate) eigenvalues $\xi_0$ and $2N$ corresponding characteristic covectors $\xi_\mu$. Strong hyperbolicity requires that these eigenvalues are all real and (in the case of degeneracy) that the algebraic multiplicity of each eigenvalue is equal to its geometric multiplicity (the dimension of the space of solutions $t_I$ to \eqref{characteristic}). 

The contracted Bianchi identity implies, for any $\xi_\mu$,
\be
\label{bianchi}
 {\cal P}_\star(\xi)^{\mu\nu\rho\sigma} \xi_\nu = 0
\ee
Hence contracting \eqref{characteristic} with $\xi_\nu$ gives
\be
\label{split_cases}
0= {\cal P}_{\rm GF}(\xi)^{\mu\nu\rho\sigma} \xi_\nu t_{\rho\sigma}= -\frac{1}{2} (\hat{g}^{\nu \gamma} \xi_\nu \xi_\gamma)( g^{\mu\beta} \tilde{P}_\beta{}^{\delta \rho \sigma} \xi_\delta t_{\rho \sigma})
\ee
So the analysis splits into two cases: either (i) $\tilde{P}_\beta{}^{\delta \rho \sigma} \xi_\delta t_{\rho \sigma}=0$ or (ii) $\hat{g}^{\nu \gamma} \xi_\nu \xi_\gamma=0$. 

{\bf Case (i)} is defined by
\be
\label{case1}
 \tilde{P}_\beta{}^{\delta \rho \sigma} \xi_\delta t_{\rho \sigma}=0
\ee
Physically, this case corresponds to a high-frequency wave with wavevector $\xi_\mu$ and polarization $t_I$ that satisfies the gauge condition \eqref{gauge_cond}. The condition \eqref{case1} implies ${\cal P}_{\rm GF}(\xi) t = 0$ and so \eqref{characteristic} reduces to 
\be
\label{case1_char}
{\cal P}_\star(\xi)^{\mu\nu\rho\sigma}t_{\rho\sigma}=0
\ee
We can divide the analysis into two subcases. 

{\bf Subcase (ia)} is defined by $g^{\gamma\delta}\xi_\gamma \xi_\delta \ne 0$. Equation \eqref{case1_char} contains a term  $-(1/2) g^{\gamma\delta}\xi_\gamma \xi_\delta t^{\mu\nu}$ and all other terms have the form $Y^{(\mu} \xi^{\nu)}$ for some $Y^{\mu}$ (depending on $t$), or are proportional to $g^{\mu\nu}$. It follows that $t_{\mu\nu}$ must have the form
\be
 t_{\mu\nu} = X_{(\mu} \xi_{\nu)} + c g_{\mu\nu}
\ee
for some $X_\mu$ and $c$. Equation \eqref{case1_char} now reduces to
$c\left( g^{\gamma \delta}\xi_\gamma \xi_\delta g^{\mu\nu} - \xi^\mu \xi^\nu \right)=0$
so $c=0$. Assuming $t_{\mu\nu} \ne 0$ (i.e. $X_\mu \ne 0$), equation \eqref{case1} now reduces to
\be
\label{tildegnull}
 \tilde{g}^{\mu\nu} \xi_\mu \xi_\nu=0
\ee
Our requirement that the null cones of $g^{\mu\nu}$ and $\tilde{g}^{\mu\nu}$ do not intersect implies that this is consistent with our starting assumption $g^{\gamma\delta}\xi_\gamma \xi_\delta \ne 0$.

Since our surfaces of constant $x^0$ are spacelike w.r.t. $\tilde{g}^{\mu\nu}$, \eqref{tildegnull} admits two real solutions $\tilde{\xi}_0^\pm$ which depend smoothly on $\xi_i$. We write the corresponding characteristic covectors as $\tilde{\xi}^\pm_\mu = (\tilde{\xi}_0^\pm,\xi_i)$, and these are null w.r.t. $\tilde{g}^{\mu\nu}$. The choice of $\pm$ corresponds to this null covector lying on either the future or past null cone of $\tilde{g}^{\mu\nu}$. Since $dx^0$ is timelike w.r.t. $\tilde{g}^{\mu\nu}$ we can distinguish these two possibilities by the sign of the non-zero quantity $(dx^0)_\mu \tilde{g}^{\mu\nu} \tilde{\xi}_\nu^\pm=  \tilde{g}^{0\nu} \tilde{\xi}_\nu^\pm$. Our convention is that $ \mp \tilde{g}^{0\nu}\tilde{\xi}_\nu^\pm>0$. 

The corresponding eigenvectors are $t_{\mu\nu} = X_{(\mu} \tilde{\xi}^\pm_{\nu)}$ where $X_\mu$ is an arbitrary covector. These are "pure gauge" eigenvectors, arising from a residual gauge freedom of \eqref{def_har}. Note that in this case we have $d$ linearly independent eigenvectors for each eigenvalue $\tilde{\xi}_0^\pm$. 

{\bf Subcase (ib)} is defined by
\be
 g^{\mu\nu} \xi_\mu \xi_\nu=0
\ee
Since surfaces of constant $x^0$ are spacelike w.r.t. $g^{\mu\nu}$ this equation admits two real solutions $\xi_0^\pm$ depending smoothly on $\xi_i$. The characteristic covector is $\xi_\mu^\pm=(\xi_0^\pm,\xi_i)$, which is null w.r.t. $g^{\mu\nu}$. We fix the signs as in case (ia) by demanding that $\mp \xi^{\pm 0}= \mp g^{0\nu}\xi_\nu^\pm>0$. The equation ${\cal P}_\star(\xi) t=0$ reduces to
\be
\label{gtransverse}
 P_\beta{}^{ \delta \rho \sigma} \xi^\pm_\delta t_{\rho \sigma} =0
\ee
 This says that the "polarization" $t_{\mu\nu}$ is transverse w.r.t. $g^{\mu\nu}$. (Note that we should really include a $\pm$ superscript on $t_{\mu\nu}$ but we suppress this to ease the notation.) However, the defining condition of case (i) gives 
\be
\label{tildegtransverse}
\tilde{P}_\beta{}^{\delta \rho \sigma} \xi^\pm_\delta t_{\rho \sigma}=0
\ee
so the polarization is also transverse w.r.t. $\tilde{g}^{\mu\nu}$. In order to solve these conditions we can introduce a basis $\{ e_0,e_1,e_{\hat{i}},\hat{i}=2,\ldots, d-1 \}$ for the tangent space such that $(e_0)^\mu = \xi^{\pm \mu}$ and $e_1^\mu \propto \xi^{\mp \mu}$, so $e_0$ and $e_1$ are both null w.r.t. $g_{\mu\nu}$. The normalization of $e_1$, and the other (spacelike) basis vectors are chosen so that
\be\label{null_basis}
 g(e_0,e_1)=1 \qquad \qquad g(e_{\hat{i}}, e_{\hat{j}}) = \delta_{\hat{i} \hat{j}}
\ee
and all other inner products of basis vectors w.r.t. $g$ vanish. Since $\xi_\mu^\pm$ depends smoothly on $\xi_i$, our basis can be chosen to depend smoothly on $\xi_i$. In such a basis, equation \eqref{gtransverse} reduces to $t_{00} = t_{0\hat{i}} = t_{\hat{i} \hat{i}} = 0$. Since the null cones of $g^{\mu\nu}$ and $\tilde{g}^{\mu\nu}$ do not intersect, it follows that $0 \ne \tilde{g}^{\mu\nu} \xi^\pm_\mu \xi^\pm_\nu = \tilde{g}^{11}$. Using this, equation \eqref{tildegtransverse} reduces to $t_{01} = 0$ and
\be
\label{tcpts}
 t_{11}  = (\tilde{g}^{11})^{-1} \tilde{g}^{\hat{i}\hat{j}} t_{\hat{i}\hat{j}} \qquad t_{1 \hat{i}} = (\tilde{g}^{11})^{-1} \tilde{g}^{1 \hat{j}} t_{\hat{i} \hat{j}}
\ee
In summary, we have shown that
\be
\label{tcpts2}
 t_{0\mu} = t_{\hat{i} \hat{i}} = 0
\ee 
and all components of $t_{\mu\nu}$ are determined (via \eqref{tcpts}) by the traceless quantity $t_{\hat{i}\hat{j}}$, which has $(1/2)d(d-3)$ independent components. Hence, for each eigenvalue $\xi_0^\pm$, $t_{\mu\nu}$ has $(1/2)d(d-3)$ independent components so the corresponding eigenspace has dimension $d(d-3)/2$. This is the number of degrees of freedom of a graviton, so these eigenvectors correspond to physical polarizations. If we choose a set of linearly independent eigenvectors for which $t_{\hat{i}\hat{j}}$ is independent of $\xi_i$ then these eigenvectors will depend smoothly on $\xi_i$. 

{\bf Case (ii)} is defined by
\be
\label{hatgnull}
 \hat{g}^{\nu \gamma} \xi_\nu \xi_\gamma=0
\ee
Since our surfaces of constant $x^0$ are spacelike w.r.t. $\hat{g}^{\mu\nu}$, it follows that this equation admits two real solutions $\hat{\xi}_0^\pm$. We write the characteristic covector as $\hat{\xi}_\mu = (\hat{\xi}_0^\pm,\xi_i)$ and fix the signs as in the previous cases by requiring that $\mp \hat{g}^{0\nu}\hat{\xi}^\pm_\nu>0$.

Recall (from \eqref{bianchi} and \eqref{split_cases}) that \eqref{hatgnull} guarantees that the contraction of \eqref{characteristic} with $\hat{\xi}^\pm_\nu$ is satisfied, i.e., $d$ components of \eqref{characteristic} are trivial. So \eqref{characteristic} is $d(d+1)/2-d$ linear equations involving the $d(d+1)/2$ components of $t_{\mu\nu}$. It follows that there must exist at least $d$ linearly independent solutions $t_{\mu\nu}$ for each eigenvalue $\hat{\xi}_0^\pm$. We can see that there exist exactly $d$ such solutions simply by counting the number of eigenvectors we have already determined. We have $2d$ eigenvectors in case (ia) ($d$ for each eigenvalue $\tilde{\xi}_0^\pm$) and $d(d-3)$ eigenvectors in case (ib) ($d(d-3)/2$ for each eigenvalue $\xi_0^\pm)$. So we have already found $d(d-1)$ eigenvectors in case (i). The total number of eigenvectors of $M(\xi_i)$ cannot exceed $2N = d(d+1)$ so we can have at most $2d$ eigenvectors in case (ii). Since we have at least $d$ eigenvectors for each eigenvalue $\hat{\xi}_0^\pm$ it follows that we must have {\it exactly} $d$ eigenvectors for each of these eigenvalues. Since these eigenvectors are associated with characteristics that are null w.r.t. $\hat{g}^{\mu\nu}$, i.e., the same as the characteristics of \eqref{Heq}, we interpret these eigenvectors as describing "gauge-condition violating" polarizations. 

We can construct these eigenvectors as follows. Since $\hat{\xi}_\mu^\pm$ is null w.r.t. $\hat{g}^{\mu\nu}$, it is spacelike w.r.t. $g^{\mu\nu}$. We now introduce a basis $\{e_0^\mu,e_1^\mu,\ldots, e_{d-1}^\mu\}$ of vectors that are orthonormal w.r.t. $g_{\mu\nu}$. We choose this basis so that $e_1^\mu$ is in the direction of the spacelike vector $\hat{\xi}^{\pm \mu}$. This orthonormal basis can be chosen so that the basis vectors depend smoothly on $\xi_i$.\footnote{To see this, start from some fixed orthonormal basis. Perform a rotation of the spatial basis vectors so that the spatial part of $\hat{\xi}^{\pm \mu}$ is in the direction $e_1^\mu$. Now perform a boost in the $1$-direction to eliminate the time component of $\hat{\xi}^{\pm \mu}$. The rotation and boost will depend smoothly on $\xi_i$ hence the new basis depends smoothly on $\xi_i$.} 

We define indices $A,B, \ldots$ to take values $0,2,3, \ldots, d-1$. As just discussed, the contraction of \eqref{characteristic} with $\hat{\xi}_\mu^\pm$ is trivial so, in our basis, only the $AB$ components of this equation are non-trivial. Furthermore, \eqref{bianchi} implies that the only non-vanishing components of ${\cal P}_\star(\hat{\xi}^\pm)^{\mu\nu\rho\sigma}$ are ${\cal P}_\star(\hat{\xi}^\pm)^{ABCD}$.

In this basis, a general symmetric tensor can be written as
\be\label{t_decomp}
 t_{\mu\nu} = \hat{\xi}^\pm_{(\mu}X_{\nu)} + t_{AB} e^A_\mu e^B_\nu
\ee
where the $1\mu$ components of $t$ are proportional to the vector $X_\mu$ of the first term. Note that this first term is in the kernel of ${\cal P}_\star(\hat{\xi}^\pm)$. 

To construct the eigenvectors, let $v^\mu$ be an arbitrary vector. Consider the equation
\begin{equation}\label{constr_viol_eq}
{\cal{P}}_\star(\hat\xi^\pm)^{ABCD} {t}_{CD}={\hat P}_\alpha{}^{\beta AB}{\hat\xi}^\pm_\beta v^\alpha
\end{equation}
We claim that this can be uniquely solved for $t_{AB}$. We will show that ${\cal P}_\star(\hat{\xi}^\pm)^{ABCD}$ has trivial kernel and is therefore invertible. So assume that $s_{AB}$ belongs to this kernel, i.e.,
\be
 {\cal{P}}_\star(\hat\xi^\pm)^{ABCD} s_{CD}=0
\ee
This implies that, for any $s_{1\mu}$,
\be\label{P_star_kernel}
 {\cal{P}}_\star(\hat\xi^\pm)^{\mu\nu\rho\sigma} s_{\rho\sigma}=0
\ee
This is the same as equation \eqref{case1_char} that we encountered in case (i) and can be solved as in subcase (ia). Using the fact that $\hat{\xi}_\mu^\pm$ is non-null w.r.t. $g^{\mu\nu}$, it follows from the tensorial structure of the equation that any such $s_{\rho\sigma}$ must have the form $s_{\rho\sigma}=c~g_{\rho\sigma}+{\hat\xi}^{\pm}_{(\rho}Y_{\sigma )}$
for some $c$ and $Y_\sigma$. Substituting this into \eqref{P_star_kernel} gives $c=0$. Hence $s_{\mu\nu}$ must be "pure gauge", i.e., the only non-trivial components are $s_{1\mu}$. In particular $s_{AB}=0$ so the kernel of ${\cal P}_\star(\hat{\xi}^\pm)^{ABCD}$ is trivial as claimed. Hence \eqref{constr_viol_eq} can be solved uniquely for $t_{AB}$. Furthermore, since the matrix on the LHS depends smoothly on $\xi_i$, it follows that the solution $t_{AB}$ will depend smoothly on $\xi_i$ and $v^\mu$. We also have 
\begin{equation}\label{constr_viol_eq2}
{\cal{P}}_\star(\hat\xi^\pm)^{\mu\nu\rho\sigma} {t}_{\rho\sigma}={\hat P}_\alpha{}^{\beta \mu\nu}{\hat\xi}^\pm_\beta v^\alpha
\end{equation}
because both sides have vanishing contraction with $\hat{\xi}_\mu^\pm$ and hence with the basis vector $e_1$. 

Next we fix $X_\mu$ by requiring that
\begin{equation}\label{harm_map}
{\tilde P}^{\mu\nu\rho\sigma}{\hat\xi^\pm}_\nu { t}_{\rho\sigma}=v^\mu
\end{equation}
This equation can be solved uniquely for $X_\mu$ in terms of $v^\mu$ and $t_{AB}$. To see this, note that the action of ${\tilde P}^{\mu\nu\rho\sigma}\hat{\xi}^\pm_\nu$ on  ${\hat\xi}^\pm_{(\rho}X_{\sigma)}$ is 
\begin{equation}
{\tilde P}^{\mu\nu\rho\sigma}{\hat\xi}^\pm_\nu{\hat\xi}^\pm_{(\rho}X_{\sigma)}=\frac12 \left({\tilde g}^{\gamma\delta}{\hat\xi}^\pm_{\gamma}{\hat\xi}^\pm_{\delta}\right) X^\mu.
\end{equation}
On the RHS, we know that $\hat{\xi}_\mu^\pm$ is non-null w.r.t. $\tilde{g}^{\mu\nu}$ because the null cones of $\tilde{g}^{\mu\nu}$ and $\hat{g}^{\mu\nu}$ do not intersect. Hence \eqref{harm_map} determines $X^\mu$ in terms of $v^\mu$ and $t_{AB}$: 
\begin{equation}\label{def_X}
X^\mu(v,t_{AB})=\frac{2}{{\tilde g}^{\gamma\delta}{\hat\xi}^\pm_\gamma{\hat\xi}^\pm_\delta}\left(v^\mu-{\tilde P}^{\mu\nu AB}{\hat\xi}^\pm_\nu ~{t}_{AB}\right)
\end{equation}
Let $t_{AB}(v)$ denote the solution of \eqref{constr_viol_eq} and let
\be
\label{tv}
 t_{\mu\nu}(v) = \hat{\xi}^\pm_{(\mu}X_{\nu)}(v,t_{AB}(v)) + t_{AB}(v) e^A_\mu e^B_\nu.
\ee
This satisfies \eqref{characteristic} because
\bea
{\cal P}(\hat \xi^{\pm})^{\mu\nu\rho\sigma}t_{\rho\sigma}(v)&=&\left({\cal P}_\star(\hat \xi^{\pm})^{\mu\nu\rho\sigma}-{\hat P}_\alpha{}^{\beta \mu\nu}{\hat\xi}^\pm_\beta {\tilde P}^{\alpha\gamma \rho\sigma}{\hat\xi}^\pm_\gamma \right)t_{\rho\sigma}(v) \nonumber \\
&=&{\hat P}_\alpha{}^{\beta \mu\nu}{\hat\xi}^\pm_\beta~v^\alpha-{\hat P}_\alpha{}^{\beta \mu\nu}{\hat\xi}^\pm_\beta {\tilde P}^{\alpha\gamma \rho\sigma}{\hat\xi}^\pm_\gamma t_{\rho\sigma}(v)=0
\eea
where we used \eqref{constr_viol_eq2} in the second equality and \eqref{harm_map} in the third. 

For every $v^\mu$ we have constructed a solution $t_{\mu\nu}$ of \eqref{characteristic} that depends smoothly on $v^\mu$ and $\xi_i$. If we choose a set of $d$ linearly independent vectors $v^\mu$ then the corresponding $t_{AB}$ are also linearly independent (using the triviality of the kernel mentioned above), and so the resulting $t_{\mu\nu}$ are linearly independent. Thus, for each eigenvalue $\hat{\xi}_0^\pm$, we have constructed a set of $d$ linearly independent eigenvectors depending smoothly on $\xi_i$.

In summary, the above calculation shows that $M(\xi_i)$ has $6$ distinct eigenvalues, namely $\xi_0^\pm$, $\tilde{\xi}_0^\pm$ and $\hat{\xi}_0^\pm$. These are all real. We have also shown that $M(\xi_i)$ has a full set of $2N$ linearly independent eigenvectors depending smoothly on $\xi_i$. Hence we have established that \eqref{def_har} is strongly hyperbolic.

\section{Lovelock theories}

\label{sec:lovelock}

\subsection{Principal symbol}

We define the modified harmonic gauge equation of motion of a Lovelock theory in exactly the same way as in GR. We start from the equation of motion in the form \eqref{eom_lovelock} and add a gauge fixing term as in \eqref{gauge_fixing} to obtain the modified harmonic gauge equation of motion in the form \eqref{def_har}, i.e.,
\be
\label{def_har_lovelock}
 {E}^{\mu\nu}_{\rm mhg} \equiv E^{\mu\nu} + \hat{P}_\alpha{}^{\beta \mu \nu} \partial_\beta H^\alpha =0
\ee
Initial data for \eqref{def_har_lovelock} consists, as in GR, of a triple $(\Sigma,h_{\mu\nu},K_{\mu\nu})$ which must satisfy the constraint equations arising from \eqref{eom_lovelock}.

The auxiliary metrics $\tilde{g}^{\mu\nu}$ and $\hat{g}^{\mu\nu}$ are chosen in the same way as in GR and we continue to raise/lower indices using $g^{\mu\nu}$ and $g_{\mu\nu}$. The argument for the propagation of the gauge condition is identical to GR (section \ref{subsec:propagation}) since it uses only the Bianchi identity for $E^{\mu\nu}$, which continues to hold in Lovelock theories.

The Lovelock equation of motion is not quasilinear, i.e., it is not linear in 2nd derivatives. We define the principal symbol as explained in appendix \ref{sec:hyperbolicity}. The result can be written as in \eqref{principal} where the gauge-fixing term is \eqref{princ_GF} and the matrix ${\cal P}_\star(\xi)$ now takes the form \cite{Choquet-Bruhat1988,Reall2014,Papallo:2017qvl}
\bea
\label{princ_star_lovelock}
 {\cal P}_\star(\xi)^{\mu\nu\rho\sigma} t_{\rho \sigma} &=&
-\frac{1}{2} g^{\gamma\delta} \xi_\gamma \xi_\delta P^{\mu\nu\rho\sigma} t_{\rho \sigma} + P_\alpha{}^{\gamma \mu\nu} \xi_\gamma g^{\alpha \beta} P_\beta{}^{ \delta \rho \sigma} \xi_\delta t_{\rho \sigma} \nonumber \\
&&-2\sum\limits_{p\geq 2} p~ k_p~ \delta_{\nu\sigma\delta\beta_3\beta_4...\beta_{2p-1}\beta_{2p}}^{\mu\rho\gamma\alpha_3\alpha_4...\alpha_{2p-1}\alpha_{2p}} t_{\rho}^{~\sigma}\xi_{\gamma}\xi^{\delta} R_{\alpha_3\alpha_4}{}^{\beta_3\beta_4}...~R_{\alpha_{2p-1}\alpha_{2p}}{}^{\beta_{2p-1}\beta_{2p}}.
\eea
In this equation, the terms in the first line arise from the Einstein tensor and the second line is the Lovelock contribution.

We can now explain what we mean by the theory being "weakly coupled". The Lovelock coupling constants $k_p$ are dimensionful. "Weakly coupled" means that the spacetime curvature is small compared to the scales defined by these constants. More precisely, it means that the terms on the second line of \eqref{princ_star_lovelock} are small compared to the terms on the first line (which don't involve the curvature). If our initial data is such that this assumption is satisfied then, by continuity, a solution of \eqref{def_har_lovelock} arising from this data will continue to be weakly coupled in a neighbourhood of $\Sigma$. However, the theory may become strongly coupled when one considers evolution over larger time intervals. For example, the theory would become strongly coupled if a curvature singularity forms. At strong coupling, well-posedness can fail \cite{Reall2014}.  

Although not quasilinear, Lovelock theories have the special property that, in any coordinate chart, the equation of motion {\it is} linear in the second derivative w.r.t. any given coordinate\footnote{\label{fnote:linear_2nd_derivative}To see this, note that the only Riemann tensor components that contain second derivatives w.r.t. $x^\alpha$ are $R_{\alpha\mu\alpha\nu}$ and components related by antisymmetry (no summation on $\alpha$). In \eqref{eom_lovelock}, the antisymmetrization over $\rho_1, \ldots, \rho_{2p}$ implies that at most one of these indices can take the value $\alpha$. Hence there are no products of second derivatives w.r.t. $x^\alpha$.} \cite{Aragone:1987jm,Choquet-Bruhat1988}, and this property is not affected by the gauge fixing term. So in a chart $x^\mu$ the modified harmonic gauge equation of motion takes the form
\be
 A^{IJ} (x,g,\partial_\mu g, \partial_0\partial_i g,  \partial_i \partial_j g) \partial_0^2 g_J = F^I(x,g,\partial_\mu g, \partial_0 \partial_i g, \partial_i \partial_j g)
\ee
where we are using the notation of section \ref{sec:GR_strong_hyperbolicity} in which indices $I, J, \ldots$ label a symmetric tensor so $g_I$ is the physical metric. $A^{IJ}$ is defined in terms of $P(\xi)$ as in \eqref{def_ABC_2}. The point is that the above equation is linear in $\partial_0^2 g_I$. A surface of constant $x^0$ is non-characteristic iff the matrix $A^{IJ}$ is invertible on that surface. Recall that in GR this is guaranteed if the surface is spacelike w.r.t. $g^{\mu\nu}$, so $\det A^{IJ} \ne 0$ on such a surface in GR. By continuity, we must also have $\det A^{IJ} \ne 0$ on a spacelike surface in Lovelock theory provided the theory is sufficiently weakly coupled. 

For non-quasilinear equations with the special property just described, the initial value problem is well-posed for initial data prescribed on a (non-characteristic) surface of constant $x^0$ provided that the equation of motion is strongly hyperbolic. (For more details, see Appendix \ref{sec:hyperbolicity}.) 
Thus to establish well-posedness we just need to demonstrate that our modified harmonic gauge Lovelock equation of motion \eqref{def_har_lovelock} is strongly hyperbolic. 

\subsection{Proof of strong hyperbolicity}

Our proof follows closely the analysis (and notation) of section \ref{sec:GR_strong_hyperbolicity}. We define the matrices $A^{IJ}$, $B^{IJ}$ and $C^{IJ}$ in terms of the principal symbol as in \eqref{def_ABC_2} and then define 
$M(\xi_i)$ with \eqref{Mdef}. We want to show that this matrix satisfies the conditions for strong hyperbolicity reviewed in section \ref{sec:GR_strong_hyperbolicity}. At weak coupling, this matrix will be close to the corresponding matrix for GR. Several steps of our argument will exploit continuity to deduce that certain features of $M(\xi_i)$ are preserved when we deform from GR to a weakly coupled Lovelock theory. 

For modified harmonic gauge GR, we showed that $M(\xi_i)$ has $6$ distinct eigenvalues. At sufficiently weak coupling, the Lovelock terms give a small deformation of the matrix $M(\xi_i)$. Since the eigenvalues of $M(\xi_i)$ depend continuously on $M(\xi_i)$, the resulting eigenvalues will fall into $6$ groups, which (following \cite{Kato1976}) we call the $\xi_0^+$-group, the $\xi_0^-$-group etc. Note that this division is possible only at weak coupling. 

At this stage we do not know that the eigenvalues of $M(\xi_i)$ are real so we view $M(\xi_i)$ as acting on the $2N$-dimensional vector space $V$ of {\it complex} vectors of the form $v=(t_{\mu\nu},t^\prime_{\mu\nu})^T$ where $t_{\mu\nu}$ and $t'_{\mu\nu}$ are symmetric. For each eigenvalue $\lambda$ we can define a "generalized eigenspace". This is the space of vectors $v$ satisfying $(M(\xi_i) - \lambda)^k v=0$ for some $k\in \{1,2,\ldots \}$. It corresponds to the sum of the Jordan blocks of $M(\xi_i)$ that have eigenvalue $\lambda$. We then define the "total generalized eigenspace" associated with the $\xi_0^+$-group as the direct sum of the generalized eigenspaces of each eigenvalue in the $\xi_0^+$-group, and similarly for the other groups \cite{Kato1976}. This gives the decomposition
\be\label{V_decomp}
 V = V^+ \oplus \tilde{V}^+\oplus \hat{V}^+ \oplus V^- \oplus \tilde{V}^- \oplus \hat{V}^-
\ee
where $V^+$ is the total generalized eigenspace associated with $\xi_0^+$-group etc. Note that these spaces depend on $\xi_i$. In GR these spaces are simply the eigenspaces associated with each eigenvalue.

We define the matrix \cite{Kato1976}
\be
\label{Pidef}
 \Pi^+ = \frac{1}{2\pi i} \int_{\Gamma^+} (M(\xi_i) - z)^{-1} dz
\ee
where $\Gamma^+$ is a circle (traversed anticlockwise) in the complex $z$-plane that encloses the point $z=\xi_0^+$ and is sufficiently small that only the eigenvalues of the $\xi_0^+$ group lie inside this circle, with all other eigenvalues lying outside this circle. The residue theorem implies that $\Pi^+:V \rightarrow V$ is a projection onto $V^+$. We can define similar projections $\tilde{\Pi}^+$ etc onto the other eigenspaces. Note that these projection operators are {\it smooth} functions of $\xi_i$, the background curvature, the Lovelock couplings etc. Note that the dimension of $V^+$ is the trace of $\Pi^+$. By continuity, this is the same for weakly coupled Lovelock theory as for GR and similarly for the dimensions of the other spaces in \eqref{V_decomp}. Hence we know that $V^\pm$ have dimension $(1/2)d(d-3)$ and the other spaces have dimension $d$.

Equation \eqref{bianchi} is a consequence of the Bianchi identity for $E^{\mu\nu}$ and therefore holds in a Lovelock theory. This implies that the argument leading to \eqref{split_cases} is valid for Lovelock theory. Thus the analysis splits into case (i) and case (ii) just as in GR. 

We start by observing that the (real) "pure gauge" eigenvectors of subcase (ia) are also eigenvectors for Lovelock theory, with the same (real) eigenvalues $\tilde{\xi}_0^\pm$. To see this, note that these eigenvectors have $t_{\mu\nu} = X_{(\mu} \tilde{\xi}^\pm_{\nu)}$, which, because of the antisymmetrization, gives a vanishing contribution to the second line of \eqref{princ_star_lovelock} (with $\xi_\mu = \tilde{\xi}_\mu^\pm$). Therefore the principal symbol acts on such $t_{\mu\nu}$ in exactly the same way as in GR so these eigenvectors are the same as in GR. 
This shows that the spaces $\tilde{V}^\pm$ are genuine eigenspaces spanned by these eigenvectors. We will discuss the Lovelock generalization of subcase (ib) (the physical eigenvectors) below. 

In case (ii) the analysis proceeds similarly to GR. This case is defined by \eqref{hatgnull}, so the (real) eigenvalues are $\hat{\xi}_0^\pm$, exactly as in GR. To construct the eigenvectors we proceed as in GR. The only step where the argument needs modifying is the demonstration that the kernel of ${\cal P}_\star^{ABCD}$ is trivial. We showed that this kernel is trivial for GR so ${\cal P}_\star^{ABCD}$ has non-vanishing determinant in GR. By continuity the determinant must remain non-zero in weakly coupled Lovelock theory. Hence the kernel is trivial for weakly coupled Lovelock theory. The rest of the argument is identical to the argument for GR. Hence one obtains $d$ real smooth eigenvectors for each eigenvalue $\hat{\xi}_0^\pm$. The spaces $\hat{V}^\pm$ are therefore genuine eigenspaces. 

It remains to discuss the "physical" eigenvalues of the $\xi_0^\pm$-groups, which correspond to subcase (ib) of the GR analysis. Generically the eigenvalues of the $\xi_0^\pm$ group will be non-degenerate, unlike the cases just discussed. Roughly speaking, this corresponds to the fact that, in a Lovelock theory, gravitational waves with different polarizations travel (in a non-trivial background) with different speeds \cite{Reall2014}. We will not attempt to construct the eigenvectors directly in this case. Instead we will construct an inner product on $V^\pm$ which we will use to build a symmetrizer for $M(\xi_i)$. 

We start by defining the matrices
\be\label{defHstar}
  H_\star^\pm = \pm \left( \begin{array}{ll} B_\star & A_\star \\ A_\star & 0 \end{array} \right)
\ee
where $A_\star$ and $B_\star$ are defined as in \eqref{def_ABC_2} but using ${\cal P}_\star$ instead of ${\cal P}$. We use these matrices to define a Hermitian form $(,)_\pm$ on $V^\pm$ as follows:
\be\label{Hstar_product}
 (v^{(1)},v^{(2)})_\pm = v^{(1)\dagger} H_{\star}^\pm v^{(2)}
\ee
 where $v^{(1)}$ and $v^{(2)}$ are in $V^\pm$. This is Hermitian because $B_\star$ and $A_\star$ are real symmetric matrices (because ${\cal P}_\star$ is real symmetric) so $H_\star^\pm$ is also real symmetric.  We will now show that this Hermitian form is positive definite and therefore defines an inner product. To do this we will show that it is positive definite for GR. By continuity (of the eigenvalues of the Hermitian form) it then follows that it is also positive definite for a weakly coupled Lovelock theory. 
 
In GR, the spaces $V^\pm$ are genuine eigenspaces with eigenvalue $\xi_0^\pm$, which implies that we have $v^{(1)} = (t^{(1)},\xi_0^\pm t^{(1)})^T$ and similarly for $v^{(2)}$. This implies that, in GR\footnote{Note that we use $*$ to denote a complex conjugate, which is different from the label $\star$ on $A_\star$ etc.}
\bea
 (v^{(1)},v^{(2)})_\pm &=&\pm t^{(1)*}_{\mu\nu} (2 \xi_0^\pm A_\star + B_\star)^{\mu\nu\rho\sigma} t^{(2)}_{\rho\sigma} \nonumber \\
 &=& \pm  t^{(1)*}_{\mu\nu} \left[-g^{0\gamma} \xi^\pm_\gamma P^{\mu\nu\rho\sigma} + P_\alpha{}^{0\mu\nu} P^{\alpha\gamma\rho \sigma} \xi^\pm_\gamma + P_\alpha{}^{\gamma \mu\nu} \xi^\pm_\gamma P^{\alpha 0 \rho \sigma} \right] t^{(2)}_{\rho\sigma} \nonumber\\
 &=& \mp \xi^{\pm 0}  t^{(1)*}_{\mu\nu} P^{\mu\nu\rho\sigma}  t^{(2)}_{\rho\sigma} 
\eea
where in the final step we used the fact that $v^{(1)}$ and $v^{(2)}$ belong to $V^\pm$ so $t^{(1)}$ and $t^{(2)}$ satisfy the condition \eqref{gtransverse}. Finally, evaluating this in the null basis we used to discuss case (ib) above, and using \eqref{tcpts} and \eqref{tcpts2} we obtain
\be
 (v^{(1)},v^{(2)})_\pm =   \mp \xi^{\pm 0} t^{(1)*}_{\hat{i}\hat{j}} t^{(2)}_{\hat{i}\hat{j}}
\ee
Since $t_{1\mu}$ is determined in terms of $t_{\hat{i}\hat{j}}$ by \eqref{tcpts}, and $\mp \xi^{\pm 0}>0$, this is indeed a positive definite Hermitian form on $V^\pm$. Having established this for GR, the result then follows for a weakly coupled Lovelock theory by continuity. 

Our next task is to show that the eigenvalues belonging to the $\xi_0^\pm$ groups are real. Consider two eigenvalues $\xi_0^{(1)}$ and $\xi_0^{(2)}$ belonging to the $\xi_0^+$-group, with corresponding eigenvectors $v^{(1)}=(t^{(1)},\xi_0^{(1)} t^{(1)})^T$ and $v^{(2)} = (t^{(2)},\xi_0^{(2)} t^{(2)})^T$. The eigenvalues and eigenvectors may be complex. Since these eigenvectors belong to case (i) they satisfy the condition
\be
\label{gttransverse}
 \tilde{P}_\beta{}^{\delta \rho \sigma} \xi_\delta^{(1)} t^{(1)}_{\rho\sigma}=0
\ee
where $\xi_\delta^{(1)} = (\xi_0^{(1)},\xi_i)$, and similarly for $t^{(2)}$. We now have
\bea
\label{Hsym}
 \left( \xi_0^{(1)} -  \xi_0^{(2)} \right) v^{(1)T} H_\star^+ v^{(2)} &=& t^{(1)}_{\mu\nu} \left[  \left( \xi_0^{(1)2} -  \xi_0^{(2)2} \right) A_\star +  \left( \xi_0^{(1)} -  \xi_0^{(2)} \right) B_\star \right]^{\mu\nu\rho\sigma} t^{(2)}_{\rho\sigma} \nonumber \\
 &=& t^{(1)}_{\mu\nu} \left[ {\cal P}_\star(\xi^{(1)}) - {\cal P}_\star(\xi^{(2)}) \right]^{\mu\nu\rho\sigma}  t^{(2)}_{\rho\sigma} \nonumber \\
 &=& t^{(2)}_{\mu\nu}  {\cal P}_\star(\xi^{(1)})^{\mu\nu\rho\sigma}  t^{(1)}_{\rho\sigma}  -t^{(1)}_{\mu\nu}  {\cal P}_\star(\xi^{(2)})^{\mu\nu\rho\sigma}  t^{(2)}_{\rho\sigma} \nonumber \\
 &=& t^{(2)}_{\mu\nu}  {\cal P}(\xi^{(1)})^{\mu\nu\rho\sigma}  t^{(1)}_{\rho\sigma}  -t^{(1)}_{\mu\nu}  {\cal P}(\xi^{(2)})^{\mu\nu\rho\sigma}  t^{(2)}_{\rho\sigma} =0
\eea
The second equality uses the definition of $A_\star$ and $B_\star$, the third equality uses the symmetry of ${\cal P}_\star$. The fourth equality follows from \eqref{gttransverse} which implies that $t^{(1)}$ is in the kernel of ${\cal P}_{\rm GF}(\xi^{(1)})$ and similarly for $t^{(2)}$. The final equality follows from \eqref{characteristic}. 

Assume that the $\xi_0^+$-group contains an eigenvalue $\xi_0$ with ${\rm Im}(\xi_0) \ne 0$ and corresponding eigenvector $v$ (belonging to $V^+$). Since $M(\xi_i)$ is real, it follows that $\xi_0^*$ is also an eigenvalue, with eigenvector $v^*$. We now set $\xi_0^{(1)} = \xi_0^*$, $v^{(1)}=v^*$, $\xi_0^{(2)} = \xi_0$ and $v^{(2)} = v$ to deduce from the above that
\be
v^\dagger H_\star^+ v = 0
\ee
i.e. $(v,v)_+=0$. But we have already seen that $(,)_+$ is positive definite in $V^+$ and so this equation implies that $v=0$, a contradiction. Hence the eigenvalues in the $\xi_0^+$-group are all real and similarly for the $\xi_0^-$ group. 

Finally we need to show that $M(\xi_i)$ is diagonalizable. Note that we have already constructed $d$ eigenvectors in each of the spaces $\tilde{V}^\pm$ and $\hat{V}^\pm$. So we just need to show that $M(\xi_i)$ is diagonalizable in $V^\pm$. To do this we need more information about the elements of $V^\pm$. Note in particular that a general element of $V^\pm$ is {\it not} an eigenvector, unlike the case of GR. 

Consider the {\it left} eigenvectors of $M(\xi_i)$. The left eigenvalues of a matrix are the same as its right eigenvalues. A simple calculation reveals that a left eigenvector with eigenvalue $\xi_0$ has the form 
\be
\label{w_def}
 w = (s_I,\xi_0 s_I) \left( \begin{array}{ll} B & A\\ A & 0 \end{array} \right)
\ee
where
\be
 s_{\mu\nu} P(\xi)^{\mu\nu\rho\sigma}=0
\ee
A family of left eigenvectors with eigenvalue $\hat{\xi}_0^\pm$ is obtained by choosing
\be
 s_{\mu\nu} = X_{(\mu} \hat{\xi}_{\nu)}^\pm
\ee
Now, from the Jordan canonical form of $M(\xi_i)$ it follows that a vector $v=(t_I,t_I')$ in any of the spaces $V^+$, $V^-$, $\tilde{V}^+$ or $\tilde{V}^-$ must be orthogonal to these left eigenvectors in the sense that
\be
0= w v = s_{\mu\nu} \left( B + \hat{\xi}_0^\pm A \right)^{\mu\nu\rho\sigma}t_{\rho \sigma} + s_{\mu\nu} A^{\mu\nu\rho\sigma}t'_{\rho\sigma}
\ee
Since $X_\mu$ is arbitrary, this implies
\be
0= \hat{\xi}^\pm_\nu \left( B + \hat{\xi}_0^\pm A \right)^{\mu\nu\rho\sigma}t_{\rho \sigma} +   \hat{\xi}^\pm_\nu  A^{\mu\nu\rho\sigma} t'_{\rho\sigma} 
\ee
This expression has to hold for both sign choices $\pm$. Note that it is quadratic in $\hat{\xi}_0^\pm$. We can eliminate $(\hat{\xi}_0^\pm)^2$ using the defining equation \eqref{hatgnull}, to obtain
\be
 \hat{\xi}_0^\pm  R^\mu + S^\mu= 0
\ee
where
\be
 R^\mu = -2 \left( \hat{g}^{00} \right)^{-1} \hat{g}^{0i} \xi_i A^{\mu0\rho\sigma} t_{\rho\sigma} + B^{\mu0 \rho \sigma} t_{\rho\sigma} + \xi_i A^{\mu i \rho \sigma} t_{\rho\sigma} + A^{\mu0 \rho\sigma} t'_{\rho\sigma}
\ee
and
\be
 S^\mu = - \left( \hat{g}^{00} \right)^{-1} \hat{g}^{ij}\xi_i \xi_j A^{\mu0 \rho \sigma} t_{\rho \sigma} + \xi_i B^{\mu i \rho \sigma}t_{\rho \sigma} + \xi_i A^{\mu i \rho \sigma} t'_{\rho \sigma}
\ee
(Recall that $\hat{g}^{00} \ne 0$ because our surfaces of constant $x^0$ are spacelike w.r.t. $g^{\mu\nu}$ and hence spacelike w.r.t. $\hat{g}^{\mu\nu}$.) Since $\hat{\xi}_0^+ \ne \hat{\xi}_0^-$, this implies that the vector $(t_I,t_I')$ must obey
\be
 R^\mu=S^\mu=0
\ee
We can simplify the expression for $R^\mu$ as follows. Equating coefficients of different powers of $\xi_0$ in \eqref{bianchi} gives
\be
 A_\star^{\mu 0 \rho \sigma} = 0 \qquad \xi_i A_\star^{\mu i \rho \sigma} + B_\star^{\mu0 \rho \sigma} = 0 \qquad \xi_i B_\star^{\mu i \rho \sigma} + C_\star^{\mu0 \rho \sigma} = 0 \qquad \xi_i C_\star^{\mu i \rho \sigma} =0 
\ee
It follows that $R^\mu$ depends only on the gauge-fixing term ${\cal P}_{\rm GF}$. A calculation now gives
\be
 R^\mu= -\frac{1}{2} \hat{g}^{00}  g^{\mu\beta} \left( \tilde{P}_\beta{}^{i \rho \sigma} \xi_i t_{\rho \sigma} + \tilde{P}_\beta{}^{0 \rho \sigma} t'_{\rho \sigma} \right)
\ee
and so $R^\mu=0$ implies
\be
\label{gentrans}
 \tilde{P}_\beta{}^{i \rho \sigma} \xi_i t_{\rho \sigma} + \tilde{P}_\beta{}^{0 \rho \sigma} t'_{\rho \sigma} =0
\ee
(As a check, note that this equation is satisfied by the {\it eigenvectors} in $V^\pm$ and $\tilde{V}^\pm$ since these have $t'_{\rho\sigma} = \xi_0 t_{\rho\sigma}$, where $\xi_0$ is the eigenvalue and $t_{\mu\nu}$ satisfies \eqref{gttransverse}.) Using this result, one can show that the gauge-fixing terms cancel out in $S^\mu$. However we will not need to consider $S^\mu$. 

We have shown that any vector in $V^\pm$ or $\tilde{V}^\pm$ must satisfy \eqref{gentrans}. Consider now the possibility that $M(\xi_i)$ is {\it not} diagonalizable in $V^+$, which means that
there is a non-trivial Jordan block associated with an eigenvalue $\xi_0$ belonging to the $\xi_0^+$ group. This implies that there is a vector $w\in V^+$ such that $(M(\xi_i) - \xi_0)^2 w = 0$ but $(M(\xi_i) - \xi_0) w \ne 0$. Hence $(M(\xi_i) - \xi_0) w$ is an eigenvector (in $V^+$) with eigenvalue $\xi_0$. Writing $w = (u_I,u'_I)^T$ this means 
\be
\label{Mu}
 \left( M(\xi_i) - \xi_0\right) \left( \begin{array}{l} u \\ u' \end{array} \right) =  \left( \begin{array}{l} t \\ \xi_0 t \end{array} \right)
\ee
for some $t_{\mu\nu}$ satisfying the characteristic condition \eqref{characteristic} and the defining condition of case (i)
\be
\label{gttransverse2}
 \tilde{P}_\alpha{}^{\gamma \rho\sigma} \xi_\gamma t_{\rho \sigma}=0
\ee
Writing out the components of \eqref{Mu} gives
\be
\label{up}
 u'_{\mu\nu} = \xi_0 u_{\mu\nu} + t_{\mu\nu}
\ee
and
\be
\label{ut}
 {\cal P}(\xi)^{\mu\nu\rho\sigma} u_{\rho \sigma} = -(2 \xi_0 A + B)^{\mu\nu\rho\sigma} t_{\rho \sigma}
\ee
Since the vector $(u_I,u'_I)$ belongs to $V^+$, it must satisfy the constraint \eqref{gentrans}. This gives
\be
\label{gentrans2}
\tilde{P}_\beta{}^{\gamma \rho\sigma} \xi_\gamma u_{\rho \sigma} + \tilde{P}_\beta{}^{0\rho\sigma} t_{\rho \sigma}=0
\ee
Now write \eqref{ut} as
\be\label{ut_2}
 {\cal P}_\star(\xi)u = -{\cal P}_{\rm GF}(\xi)u  - (2 \xi_0 A_{\rm GF} + B_{\rm GF}) t - (2 \xi_0 A_\star + B_\star) t.
\ee
Using \eqref{gentrans2} and \eqref{gttransverse2} we find that
\bea\label{gf_cancel}
-\left[{\cal P}_{\rm GF}(\xi)u  + (2 \xi_0 A_{\rm GF} + B_{\rm GF}) t\right]^{\mu\nu} &=& {\hat{  P}}_{\alpha}^{~\delta\mu\nu}\xi_\delta{\tilde{  P}}^{\alpha\gamma\rho\sigma}\xi_\gamma u_{\rho\sigma}+{\hat{  P}}_{\alpha}^{~\delta\mu\nu}\xi_\delta{\tilde{  P}}^{\alpha 0\rho\sigma} t_{\rho\sigma}+{\hat{  P}}_{\alpha}^{~0\mu\nu}{\tilde{  P}}^{\alpha\gamma\rho\sigma}\xi_\gamma t_{\rho\sigma} \nonumber \\
&=& {\hat{  P}}_{\alpha}^{~0\mu\nu}{\tilde{  P}}^{\alpha\gamma\rho\sigma}\xi_\gamma t_{\rho\sigma}=0,
\eea
i.e. the gauge-fixing terms all cancel on the RHS of \eqref{ut_2}, leaving
\be
 {\cal P}_\star(\xi)^{\mu\nu\rho\sigma}u_{\rho\sigma} = - (2 \xi_0 A_\star + B_\star)^{\mu\nu\rho\sigma} t_{\rho\sigma}.
\ee
Now contract this equation with $t^*_{\mu\nu}$ (the complex conjugate of $t_{\mu\nu}$). Since ${\cal P}_\star$ is symmetric we can write the LHS as $u{\cal P}_\star(\xi) t^*$ and this vanishes because $P_\star(\xi) t^* = (P_\star(\xi) t)^* = (P(\xi) t)^* = 0$ using the fact that $t$ satisfies \eqref{gttransverse2}. So we are left with
\be
 0 = t^*_{\mu\nu} (2 \xi_0 A_\star + B_\star)^{\mu\nu\rho\sigma} t_{\rho\sigma} = (v,v)_+
\ee
where $v=(t_I,\xi_0 t_I)$ is the eigenvector. But we showed earlier that $(,)_+$ is positive definite. So we must have $v=0$, which is a contradiction. Hence our assumption that there exists a non-trivial Jordan block must be false. Therefore $M(\xi_i)$ must be diagonalizable within $V^+$. A similar argument demonstrates that $M(\xi_i)$ is diagonalizable within $V^-$. 

We have now proved that $M(\xi_i)$ is diagonalizable with real eigenvalues. Since $M(\xi_i)$ is real, this implies that we can choose the eigenvectors to be real. Our final task is to construct a symmetrizer. Note that the eigenvalues associated with $V^\pm$ are generically distinct and have non-trivial dependence on the Riemann tensor and $\xi_i$ \cite{Reall2014}. However, as $\xi_i$ is varied, the eigenvalues might cross and if this happens then the eigenvectors might not be smooth functions of $\xi_i$ \cite{Kato1976}. This means that the standard choice of a symmetrizer in the subspaces $V^\pm$ might not be smooth in $\xi_i$, so the definition of strong hyperbolicity would not be satisfied. However, one can easily overcome this difficulty. Given two eigenvectors $v^{(1)}$ and $v^{(2)}$ in $V^\pm$ with respective eigenvalues $\xi_0^{(1)}$ and $\xi_0^{(2)}$ we have
\be
v^{(1)T}\left(M^T H_\star^\pm-H_\star^\pm M\right)v^{(2)}=\left( \xi_0^{(1)} -  \xi_0^{(2)} \right) v^{(1)T} H_\star^\pm v^{(2)}=0
\ee
where the final equality is \eqref{Hsym}. Since the eigenvectors form a basis for $V^\pm$, it follows that $H_\star^\pm$ is a symmetrizer for $M(\xi_i)$ within $V^\pm$. Crucially, $H_\star^\pm$ depends smoothly on $\xi_i$. We now construct a symmetrizer for $M(\xi_i)$ within $V$ as a block diagonal matrix where the blocks associated with $\tilde{V}^\pm$ and $\hat{V}^\pm$ are constructed from the (smooth) eigenvectors in the usual way, and the blocks associated with $V^\pm$ are equal to $H_\star^\pm$. More explicitly, let $\{v_1^\pm, ..., v_{d(d-3)/2}^\pm\}$ be a smooth basis for $V^\pm$ and let $\{{\tilde v}_1^\pm, ..., {\tilde v}_d^\pm\}$ and $\{{\hat v}_1^\pm, ..., {\hat v}_d^\pm\}$ denote the smooth eigenvectors (constructed above) in ${\tilde V}^\pm$ and ${\hat V}^\pm$ respectively. Furthermore, let $S$ be the matrix whose columns are these (real) basis vectors. Then $M(\xi_i)$ can be written as
\be
M(\xi_i)=S\left(\begin{array}{cccccc}
    \Xi^+ & 0&0&0&0&0  \\
    0 & {\tilde \xi}_0^+ I_d & 0&0&0&0 \\
    0&0& {\hat \xi}_0^+ I_d &0&0&0 \\
    0&0&0&\Xi^- & 0&0  \\
    0&0&0&0 & {\tilde \xi}_0^- I_d & 0 \\
    0&0&0&0&0& {\hat \xi}_0^- I_d  \\
\end{array}\right) S^{-1}
\ee
where $\Xi^\pm$ are $d(d-3)/2\times d(d-3)/2$ matrices and $I_d$ is the $d\times d$ unit matrix. Our proposal for the symmetrizer can also be written in a decomposed form as
\be\label{symm_lovelock}
K(\xi_i)=\left(S^{-1}\right)^T \left(\begin{array}{cccccc}
    {\cal H}_\star^+ & 0&0&0&0&0  \\
    0 &  I_d & 0&0&0&0 \\
    0&0&  I_d &0&0&0 \\
    0&0&0&{\cal H}_\star^- & 0&0  \\
    0&0&0&0 &  I_d & 0 \\
    0&0&0&0&0&  I_d  \\
\end{array}\right) S^{-1}
\ee
where ${\cal H}_\star^\pm$ are the $d(d-3)/2\times d(d-3)/2$ matrices whose elements are $(\hat A, \hat B=1,2,...,d(d-3)/2)$
\be
\left({\cal H}_\star^\pm\right)_{\hat A \hat B}=\left(v_{\hat A}^\pm\right)^T H_\star^\pm v_{\hat B}^\pm
\ee
The matrix given by \eqref{symm_lovelock} then clearly satisfies \eqref{sym_def} and therefore it is indeed a symmetrizer. Moreover, as explained above, it depends smoothly on $\xi_i$. This concludes the proof of strong hyperbolicity.

\section{Horndeski theories}

\label{sec:horndeski}

In this section, we will analyze the hyperbolicity of the equations of motion of weakly coupled Horndeski theories in modified harmonic gauge. Our discussion uses similar ideas to the Lovelock case so to avoid repetition, we will merely point out the differences.

\subsection{Principal symbol}

We will write the action of a Horndeski theory in the form 
	\begin{equation}\label{horndeski_action}
	S=\frac{1}{16\pi G}\int \mathrm{d}^4 x\sqrt{-g} \left(\mathcal{L}_1+\mathcal{L}_2+\mathcal{L}_3+\mathcal{L}_4+\mathcal{L}_5\right)
	\end{equation}
with
	\begin{align*}
	\mathcal{L}_1=&{}~R+X -V(\phi)\\
	\mathcal{L}_2=&{}~G_2(\phi,X)\\
	\mathcal{L}_3=&{}~G_3(\phi,X)\Box \phi\\
	\mathcal{L}_4=&{}~G_4(\phi,X){R}+\partial_X G_4(\phi,X)\delta_{\rho\sigma}^{\mu\nu}\nabla_\mu\nabla^\rho\phi\nabla_\nu\nabla^\sigma\phi\\
	\mathcal{L}_5=&{}~G_5(\phi,X){G}_{\mu\nu}\nabla^\mu\nabla^\nu\phi-\frac16 \partial_X G_5(\phi,X)\delta_{\alpha\beta\gamma}^{\mu\nu\rho}\nabla_\mu\nabla^\alpha\phi\nabla_\nu\nabla^\beta\phi\nabla_\rho\nabla^\gamma\phi
	\end{align*}
where
\be
 X\equiv -\frac12 (\partial\phi)^2
\ee
\noindent
and $G_i$ $(i=2,3,4,5)$ are arbitrary coupling functions. Note that ${\cal L}_1$ is the action of Einstein gravity minimally coupled to a scalar field with potential $V$, we will refer to this as {\it Einstein-scalar field theory}. We could absorb the terms in ${\cal L}_1$ into other terms; the reason we do not do this is that we want to regard a weakly coupled Horndeski theory as a small deformation of GR. 

Variation of \eqref{horndeski_action} w.r.t. the metric and scalar field yields the equations of motion
\be
\label{horndeski_eofm}
E^{\mu\nu} \equiv -16\pi G\frac{1}{\sqrt{-g}}\frac{\delta S}{\delta g_{\mu\nu}}=0, \qquad \qquad
E_\phi \equiv -16\pi G\frac{1}{\sqrt{-g}}\frac{\delta S}{\delta \phi}=0 
\ee
The explicit form of these equations for the most general Horndeski theory can be found in e.g. Appendix A of \cite{Papallo:2017ddx}. 

Our modified harmonic gauge equations of motion are defined in exactly the same way as in GR and for Lovelock theories. We introduce the two auxiliary (inverse) metrics $\tilde{g}^{\mu\nu}$ and $\hat{g}^{\mu\nu}$ satisfying the same conditions as in section \ref{subsec:GR_eom}. The gravitational equation of motion is modified as in equation (\ref{gauge_fixing}) and the scalar equation is left unchanged, so the equations of motion are 
\be
{E}^{\mu\nu}_{\rm mhg} \equiv  E^{\mu\nu}+{\hat P}_\alpha{}^{\beta\mu\nu}\partial_\beta H^\alpha=0 \label{Horndeski_eom_1} 
\ee
and
\be
E_\phi =0 . \label{Horndeski_eom_2}
\ee
Initial data for Horndeski theories consists of a quintuple $(\Sigma,h_{\mu\nu},K_{\mu\nu},\Phi,\Psi)$ where $h_{\mu\nu}$ and $K_{\mu\nu}$ correspond to the induced metric and extrinsic curvature tensor of $\Sigma$, and $\Phi$, $\Psi$ to the values of $\phi$ and its normal derivative on $\Sigma$. The equations $E^{\mu\nu}n_\nu=0$ contain no second time derivatives of $\phi$ and $g$ so they are constraint equations that the initial data must satisfy.

The diffeomorphism invariance of the action implies that, for any $g_{\mu\nu}$ and $\phi$, we have the generalized Bianchi identity
\be\label{scalar_bianchi}
\nabla^\mu E_{\mu\nu}-E_\phi\nabla_\nu\phi=0.
\ee
Since there are no gauge fixing terms in the scalar equation of motion, this identity implies that the equation describing the propagation of the gauge condition $H^\mu$ remains the same as in GR, i.e., the deformed harmonic gauge equations of motion implies that $H^\mu$ satisfies \eqref{Heq}. Thus if $H^\mu$ and its normal derivative vanish on $\Sigma$ then $H^\mu$ will vanish throughout $\hat{D}(\Sigma)$, as in section \ref{subsec:propagation}.

To construct initial data for \eqref{Horndeski_eom_1}, \eqref{Horndeski_eom_2} we proceed as in section \ref{subsec:propagation}, writing the metric in terms of the lapse and shift to ensure that the surface $x^0=0$ is spacelike w.r.t. $g^{\mu\nu}$. The time derivative of the lapse and shift are chosen to ensure that $H^\mu=0$ at $x^0=0$ and the constraint equations then imply that the derivative of $H^\mu$ vanishes at $x^0=0$ as in section \ref{subsec:propagation}. Hence $H^\mu$ vanishes throughout $\hat{D}(\Sigma)$ and so the resulting solution of 
\eqref{Horndeski_eom_1} and \eqref{Horndeski_eom_2} is also a solution of the original Horndeski equations of motion in $\hat{D}(\Sigma)$ and hence also in $D(\Sigma)$. 

A general Horndeski theory is not quasilinear. We define the principal symbol by varying the equation of motion w.r.t. the second derivatives of the fields, as explained in Appendix \ref{sec:hyperbolicity}. The principal symbol acts on a vector of the form $T_I\equiv (t_{\mu\nu},\psi)^T$ where $t_{\mu\nu}$ is symmetric. The space of such vectors is $11$-dimensional so indices $I,J,\ldots$ take values from $1$ to $11$. We label the different blocks of the principal symbol as follows:
\be
\label{princ_horndeski}
 {\cal P}(\xi)^{IJ} ={\cal P}^{IJ\gamma\delta}\xi_\gamma\xi_\delta=\left(\begin{array}{cc}
  {\cal P}_{gg}(\xi)^{\mu\nu\rho\sigma} & {\cal P}_{g\phi }(\xi)^{\mu\nu} \\
  {\cal P}_{\phi g }(\xi)^{\rho\sigma} & {\cal P}_{\phi \phi }(\xi)
 \end{array}\right).
\ee
In other words:
\be
\label{PT}
 {\cal P}(\xi)^{IJ} T_J=\left(\begin{array}{c}
  {\cal P}_{gg}(\xi)^{\mu\nu\rho\sigma}t_{\rho\sigma} + {\cal P}_{g\phi }(\xi)^{\mu\nu}\psi \\
  {\cal P}_{\phi g }(\xi)^{\rho\sigma}t_{\rho\sigma} + {\cal P}_{\phi \phi }(\xi)\psi
 \end{array}\right).
\ee
We decompose the principal symbol as in \eqref{principal} into a part ${\cal P}_\star$ coming from \eqref{horndeski_eofm} and a part ${\cal P}_{\rm GF}$ coming from the gauge fixing term. The former can be written
\be
\label{princ_star_horndeski}
 {\cal P}_\star(\xi)^{IJ}=\left(\begin{array}{cc}
  {\cal P}_{gg\star}(\xi)^{\mu\nu\rho\sigma} & {\cal P}_{g\phi \star}(\xi)^{\mu\nu} \\
  {\cal P}_{\phi g \star}(\xi)^{\rho\sigma}& {\cal P}_{\phi \phi \star}(\xi)
 \end{array}\right).
\ee
where the $11\times 11$ matrix $\mathcal{P}_\star(\xi)^{IJ}$ is {\it symmetric} because the equations of motion (without gauge-fixing) are obtained from an action \cite{Papallo:2017qvl}. In particular we have
\be
{\cal P}_{g\phi \star}(\xi)^{\mu\nu} = {\cal P}_{\phi g \star}(\xi)^{\mu\nu}
\ee
The contribution of the gauge-fixing term to the principal symbol is 
\be\label{princ_horndeski_gf}
{\cal P}_{\rm GF}(\xi)^{IJ} =\left(\begin{array}{cc}
   - {\hat P}_\alpha{}^{\gamma \mu\nu} \xi_\gamma g^{\alpha \beta} {\tilde P}_\beta{}^{ \delta \rho \sigma} \xi_\delta   & 0 \\
  0 & 0
 \end{array}\right)
\ee
It is useful to split $\cal{P}_\star$ into the sum of two terms corresponding to the contributions from the Einstein-scalar field Lagrangian ${\cal L}_1$ and the Horndeski terms ${\cal L}_i$ with $i \ge 2$ respectively:
\be
\label{horndeski_split}
 {\cal P}_\star(\xi)^{IJ} = {\cal P}_{\star}^\text{Esf}(\xi)^{IJ}+\delta {\cal P}_\star(\xi)^{IJ}
\ee
where the Einstein-scalar-field part is
\be
\label{esf_symbol}
{\cal P}_{\star}^\text{Esf}(\xi)^{IJ}=\left(\begin{array}{cc}
 -\frac{1}{2} g^{\gamma\delta} \xi_\gamma \xi_\delta P^{\mu\nu\rho\sigma}  + P_\alpha{}^{\gamma \mu\nu} \xi_\gamma g^{\alpha \beta} P_\beta{}^{ \delta \rho \sigma} \xi_\delta   & 0 \\
  0 & -g^{\gamma\delta}\xi_\gamma\xi_\delta
 \end{array}\right).
\ee
The form of the Horndeski terms $\delta {\cal P}_\star(\xi)^{IJ}$ can be found in Appendix B of \cite{Papallo:2017ddx}. As in the case of Lovelock theories, we define "weak coupling" to mean that $\delta {\cal P}_\star(\xi)^{IJ}$ is small compared to ${\cal P}_{\star}^\text{Esf}(\xi)^{IJ}$. This will be the case if the Riemann tensor and first and second derivatives of the scalar field are small compared to any length scales defined by the (dimensionful) Horndeski coupling functions $G_i(\phi,X)$ and their derivatives w.r.t. $X$. As in the Lovelock case, if the initial data is chosen so that the theory is weakly coupled initially then, by continuity, the resulting solution will remain weakly coupled at least for a small time. But the theory may become strongly coupled over a longer time, in which case even weak hyperbolicity of the equations of motion may fail \cite{Papallo:2017qvl,Ripley:2019hxt,Ripley:2019irj,Ripley:2019aqj}. 

The equations of motion of a generic Horndeski theory are not quasilinear but they have the same special structure as Lovelock theories, i.e., in any chart, they are linear in second derivatives w.r.t. any given coordinate\footnote{This can be seen from the equations of motion written out in Appendix A of Ref. \cite{Papallo:2017ddx}. Second derivatives w.r.t. $x^\alpha$ appear only in the Riemann tensor component $R_{\alpha \mu \alpha \nu}$ (or components related by antisymmetry) and in $\nabla_\alpha \nabla_\alpha \phi$. Non-quasilinear terms all have antisymmetrizations which prevent two indices $c_i$ (in the equations of \cite{Papallo:2017ddx}) being equal to $\alpha$ and hence products of second derivative w.r.t. $x^\alpha$ do not appear.}, and this property is not affected by the gauge fixing term. Hence, in a chart $x^\mu=(x^0,x^i)$, the modified harmonic gauge equation of motion can be written as
\be
A^{IJ}(x,u,\partial_\mu u, \partial_0\partial_i u,\partial_i\partial_j u)\partial_0^2u_J=F^{I}(x,u,\partial_\mu u, \partial_0\partial_i u,\partial_i\partial_j u)
\ee
with $u_I=(g_{\mu\nu},\phi)$ and $A^{IJ}$ is defined in terms of ${\cal P}^{IJ}$ as in \eqref{def_ABC_2}. So the equations are linear in the second time derivatives of the fields. In Einstein-scalar-field theory, the matrix $A^{IJ}$ is invertible on surfaces of constant $x^0$ provided such surfaces are spacelike, i.e., spacelike surfaces are non-characteristic. By continuity, a spacelike hypersurface remain non-characteristic for a sufficiently weakly coupled Horndeski theory. Therefore, if we can show that the system \eqref{Horndeski_eom_1}-\eqref{Horndeski_eom_2} is strongly hyperbolic then the results reviewed in Appendix \ref{sec:hyperbolicity} will apply and local well-posedness of the initial value problem for weakly coupled Horndeski theories would be established.

\subsection{Proof of strong hyperbolicity}

Using a coordinate system $x^\mu=(x^0,x^i)$, the $11\times 11$ matrices $A^{IJ}$, $B^{IJ}$, $C^{IJ}$ are defined by (\ref{def_ABC_2}). To analyze the hyperbolicity of the equations of motion \eqref{Horndeski_eom_1}-\eqref{Horndeski_eom_2}, we must study the eigenvalue problem of the $22\times 22$ matrix $M(\xi_i)$ defined by equation \eqref{Mdef}, where $\xi_i$ is a real covector with unit norm w.r.t. an arbitrary smooth (inverse) Riemannian metric $G^{ij}$ defined on surfaces of constant $x^0$. This matrix acts on the $22$-dimensional vector space $V$ of complex vectors of the form $v=(T_I,T_I^\prime)^T$ with $T_I=(t_{\mu\nu},\psi)^T$, $T_I^\prime=(t^\prime_{\mu\nu},\psi^\prime)^T$. An eigenvector corresponding to an eigenvalue $\xi_0$ has the form $(T_I,\xi_0 T_I)^T$ where $T_I$ satisfies
\be\label{characteristic_horndeski}
{\cal P}(\xi)^{IJ}T_J=0
\ee
with $\xi_\mu=(\xi_0,\xi_i)$, i.e., $\xi_\mu$ is characteristic with associated "polarization" $T_I$.

We start by considering Einstein-scalar field theory. If $\xi_0$ is an eigenvalue for vacuum GR, with eigenvector $(t_{\mu\nu},\xi_0 t_{\mu\nu})^T$ then $\xi_0$ is also an eigenvalue for Einstein-scalar-field theory with eigenvector $(T_I,\xi_0 T_I)^T$ where $T_I = (t_{\mu\nu},0)^T$ (i.e. $T_I$ has $\psi=0$). This gives us "pure gauge" eigenvalues $\tilde{\xi}_0^\pm$ and "gauge-condition violating" eigenvalues $\hat{\xi}_0^\pm$, each with $4$ eigenvectors, and "physical" eigenvalues $\xi_0^\pm$, each with $2$ eigenvectors. A further physical eigenvector with eigenvalue $\xi_0^\pm$ is obtained by setting $t_{\mu\nu}=0$ and $\psi=1$. Thus for each physical eigenvalue $\xi_0^\pm$ there are $3$ eigenvectors, corresponding to $2$ graviton polarizations and $1$ scalar field polarization. The eigenvalues are all real, the total number of eigenvectors is $22$, and the eigenvectors depend smoothly on $\xi_i$. Hence our modified harmonic gauge formulation of Einstein-scalar field theory is strongly hyperbolic. 

Now we consider a weakly coupled Horndeski theory. Just as for a weakly coupled Lovelock theory, continuity of the eigenvalues implies that the eigenvalues of $M(\xi_i)$ for weakly coupled Horndeski theories can be split into $6$ groups. The decomposition
\be
 V = V^+ \oplus \tilde{V}^+\oplus \hat{V}^+ \oplus V^- \oplus \tilde{V}^- \oplus \hat{V}^-
\ee
into total generalized eigenspaces and the definition of the corresponding projection matrices \eqref{Pidef} (that depend smoothly on $\xi_i$, the background fields and the Horndeski couplings) is the same as for the Lovelock case. The counting of eigenvectors for Einstein-scalar field theory implies that the spaces $\tilde{V}^\pm$ and $\hat{V}^\pm$ are $4$-dimensional whereas $V^\pm$ are $3$-dimensional. 

Analogously to equation \eqref{bianchi}, diffeomorphism invariance of the action implies that \cite{Papallo:2017qvl}
\be
\label{princ_gauge}
{\cal P}_{gg\star}(\xi)^{\mu\nu\rho\sigma}\xi_\nu=0,\qquad {\cal P}_{g\phi\star}(\xi)^{\mu\nu}\xi_\nu=0.
\ee
The characteristic equation is \eqref{characteristic_horndeski}. Writing the LHS as in \eqref{PT}, taking the contraction of the first row with $\xi_\nu$, and using \eqref{princ_gauge}, we obtain equation \eqref{split_cases}. Hence the analysis splits into case (i) and case (ii) just as for vacuum GR and for Lovelock theories. For Einstein-scalar field theory we can split case (i) into subcases (ia) and (ib), as in GR, where the physical eigenvectors with $t_{\mu\nu}=0$ are included in subcase (ib). 

In Einstein-scalar field theory, subcase (ia) gives the "pure gauge" eigenvectors with eigenvalues $\tilde{\xi}_0^\pm$ and $T_I= (\tilde{\xi}^\pm_{(\mu}X_{\nu)},0)^T$ for any $X_\nu$. These are also eigenvectors for a Horndeski theory, with the same (real) eigenvalues $\tilde{\xi}_0^\pm$. So the spaces $\tilde{V}^\pm$ are genuine eigenspaces spanned by these eigenvectors. In Einstein-scalar field theory, subcase (ib) gives the "physical" eigenvectors with eigenvalues $\xi_0^\pm$. We will discuss the Horndeski generalization of these below.

Case (ii) is defined by \eqref{hatgnull}, so the (real) eigenvalues are $\hat{\xi}_0^\pm$ as in Einstein-scalar field theory. The corresponding "gauge condition violating" eigenvectors can be constructed similarly to the Lovelock case. One can introduce a smooth orthonormal (w.r.t. $g_{\mu\nu}$) basis adapted to $\hat\xi^\pm_\mu$ (see the paragraphs above \eqref{t_decomp}) with indices $A,B \ldots$ labelling directions orthogonal to $\hat\xi^\pm_\mu$, which is spacelike w.r.t. $g^{\mu\nu}$. Equation \eqref{princ_gauge} implies that the only non-vanishing components of ${\cal P}_{gg\star}(\hat{\xi}^\pm)^{\mu\nu\rho\sigma}$ and ${\cal P}_{g\phi\star}(\hat{\xi}^\pm)^{\mu\nu}$ are the $ABCD$ and $AB$ components respectively. 

We then fix a vector $v^\mu$ and consider the equation
\be
\label{tpsiv}
\left(\begin{array}{cc}
  {\cal P}_{gg\star}(\hat{\xi}^\pm)^{ABCD} & {\cal P}_{g\phi \star}(\hat{\xi}^\pm)^{AB} \\
  {\cal P}_{\phi g \star}(\hat{\xi}^\pm)^{CD}& {\cal P}_{\phi \phi \star}(\hat{\xi}^\pm)\end{array}\right) \left( \begin{array}{c} t_{CD} \\ \psi \end{array} \right) = \left( \begin{array}{c} {\hat P}_\alpha{}^{\beta AB}{\hat\xi}^\pm_\beta v^\alpha \\ 0 \end{array} \right)
\ee
We will show that the matrix on the LHS is invertible in a weakly coupled Horndeski theory. Consider first the case of Einstein-scalar field theory. In this case, the matrix on the LHS is block diagonal and $(s_{AB},\chi)$ belongs to the kernel iff
\bea
{\cal P}_{gg \star}(\hat\xi^\pm)^{ABCD}s_{CD}&=0 \nonumber \\
g^{\gamma\delta}{\hat\xi^\pm}_\gamma{\hat\xi^\pm}_\delta\chi=0.
\eea
Since $g^{\gamma\delta}{\hat\xi^\pm}_\gamma{\hat\xi^\pm}_\delta\neq 0$, we have $\chi=0$ and the argument used in the vacuum GR case establishes that $s_{AB}=0$, so the kernel is trivial in Einstein-scalar field theory. Thus the matrix on the LHS of \eqref{tpsiv} has non-vanishing determinant in Einstein-scalar field theory. By continuity, its determinant must be non-zero for a weakly coupled Horndeski theory. Hence this matrix is invertible. So, for each $v^\mu$, this equation uniquely defines $(t_{AB}(v),\psi(v))$. This will depend smoothly on $v^\mu$. It also depends smoothly on $\xi_i$ because the matrix on the LHS, and the RHS of \eqref{tpsiv} depend smoothly on $\xi_i$. 

The rest of the argument proceeds as in GR: define $X_\mu(v)$ as in \eqref{def_X} and $t_{\mu\nu}(v)$ as in \eqref{tv}. Let $T_I(v)=(t_{\mu\nu}(v),\psi(v))^T$. We then have ${\cal P}^{IJ}(\hat{\xi}^\pm) T_J(v) = 0$. Thus for each $v^\mu$ we have constructed an eigenvector. Letting $v^\mu$ run over a basis of $4$ linearly independent vectors gives us $4$ linearly independent eigenvectors. Thus we have proved that $\hat{V}^\pm$ are genuine eigenspaces. The eigenvectors depend smoothly on $\xi_i$.

It remains to show that the physical spaces $V^\pm$ are genuine eigenspaces. 
Following the argument we used for Lovelock theories, we define the matrix $H_\star$ by equation \eqref{defHstar} and the corresponding Hermitian form on $V^\pm$ by \eqref{Hstar_product}. In Einstein-scalar field theory, $V^\pm$ are genuine eigenspaces (this is subcase (ib)) and we can use the null basis introduced above equation \eqref{null_basis} to show that this Hermitian form is
\be
 (v^{(1)},v^{(2)})_\pm =   \mp \xi^{\pm 0}\left( t^{(1)*}_{\hat{i}\hat{j}} t^{(2)}_{\hat{i}\hat{j}}+2\psi^* \psi\right).
\ee
This is positive definite. By continuity, it remains positive definite for a weakly coupled Horndeski theory. Hence we have shown that our Hermitian form defines an inner product on $V^\pm$. 

Just as for a Lovelock theory, the defining equation of case (i), i.e. \eqref{gttransverse}, and the symmetry of $\cal{P}_\star$ imply the identity \be\label{Hsym2}
 \left( \xi_0^{(1)} -  \xi_0^{(2)} \right) v^{(1)T} H_\star^+ v^{(2)}  =0
\ee
(c.f. \eqref{Hsym}) for two eigenvectors $v^{(1)}=(T^{(1)},\xi_0^{(1)} T^{(1)})^T$ and $v^{(2)} = (T^{(2)},\xi_0^{(2)} T^{(2)})^T$ belonging to $V^\pm$ with respective eigenvalues $\xi_0^{(1)}$ and $\xi_0^{(2)}$. We then follow the argument used for Lovelock theory to conclude that positive definiteness of the inner product on $V^\pm$ ensures that the eigenvalues in the $\xi_0^\pm$-groups are real, as in a weakly coupled Lovelock theory. 

The final step is to show that $M(\xi_i)$ is diagonalizable on $V^\pm$. Again we follow the argument used for Lovelock theories and consider the left eigenvectors of $M(\xi_i)$. The left eigenvectors corresponding to the eigenvalue $\hat{\xi}_0^\pm$ have the form \eqref{w_def} where $s_I=(\hat{\xi}_{(\mu}^\pm X_{\nu)},0)^T$, for arbitrary $X_\mu$. From the Jordan decomposition of $M(\xi_i)$, the subspaces $V^+$, $V^-$, $\tilde V^+$ and $\tilde{V}^-$ must be orthogonal to these eigenvectors for any $X_\mu$ and both choices of sign in $\hat\xi_\mu^\pm$. For $v=(T_I,T_I')^T$ in one of these subspaces, this implies that\footnote{
Recall that indices $I,J,\ldots$ label vectors of the form $T_I = (t_{\mu\nu},\psi)^T$ where $t_{\mu\nu}$ is symmetric. $A^{\mu 0 J}$ means the $I=(\mu 0)$ component of $A^{IJ}$, i.e., the component corresponding to $t_{\mu 0}$.} 
\be
 R^\mu \equiv -2 \left( \hat{g}^{00} \right)^{-1} \hat{g}^{0i} \xi_i A^{\mu0J} T_{J} + B^{\mu0 J} T_{J} + \xi_i A^{\mu i J} T_{J} + A^{\mu0 J} T'_{J}=0
\ee
and
\be
 S^\mu \equiv - \left( \hat{g}^{00} \right)^{-1} \hat{g}^{ij}\xi_i \xi_j A^{\mu0 J} T_{J} + \xi_i B^{\mu i J}T_{J} + \xi_i A^{\mu i J} T'_{J}=0
\ee
Similarly to the Lovelock case, writing out equation \eqref{princ_gauge} in terms of the coefficients $A_\star^{IJ}$, $B_\star^{IJ}$ and $C_\star^{IJ}$ gives
\be
 A_\star^{\mu 0 I} = 0 \qquad \xi_i A_\star^{\mu i I} + B_\star^{\mu0 I} = 0 \qquad \xi_i B_\star^{\mu i I} + C_\star^{\mu0 I} = 0 \qquad \xi_i C_\star^{\mu i I} =0 
\ee
This implies that $R^\mu$ depends only on the principal symbol of the gauge-fixing terms, given in \eqref{princ_horndeski_gf}. Writing $T_I = (t_{\mu\nu},\psi)^T$ and $T'_I = (t'_{\mu\nu},\psi')^T$, the expression for $R^\mu$ reduces to
\be
 R^\mu= -\frac{1}{2} \hat{g}^{00}  g^{\mu\beta} \left( \tilde{P}_\beta{}^{i \rho \sigma} \xi_i t_{\rho \sigma} + \tilde{P}_\beta{}^{0 \rho \sigma} t'_{\rho \sigma} \right).
\ee
Therefore any vector in $V^\pm$ (or $\tilde V^\pm$) must satisfy
\be
\label{gentrans_horndeski}
 \tilde{P}_\beta{}^{i \rho \sigma} \xi_i t_{\rho \sigma} + \tilde{P}_\beta{}^{0 \rho \sigma} t'_{\rho \sigma} =0.
\ee

The proof of the diagonalizability of $M(\xi_i)$ in $V^\pm$ is the same as for a weakly coupled Lovelock theory. Assume that there exists a non-trivial Jordan block in $V^+$ with corresponding eigenvalue $\xi_0$. Then there must be a vector $w\equiv (U_I,U_I^\prime)^T\in V^+$ such that $(M(\xi_i)-\xi_0)w\neq 0$ is an eigenvector $v\equiv(T_I,\xi_0 T_I)\in V^+$ with eigenvalue $\xi_0$. This is equivalent to the equations
\be\label{1st_row}
 U'_I = \xi_0 U_I + T_I
\ee
and
\be\label{2nd_row}
 {\cal P}(\xi)^{IJ} U_J = -(2 \xi_0 A + B)^{IJ} T_{J}.
\ee
where $T_I$ satisfies \eqref{characteristic_horndeski} and if we write $T_I = (t_{\mu\nu},\psi)$ then $t_{\mu\nu}$ satisfies \eqref{case1} (the defining condition of case (i)).

Decomposing \eqref{2nd_row} into the contributions of $\cal P_\star$ and ${\cal P}_\text{GF}$ as in \eqref{ut_2}, it can be shown that the gauge fixing terms cancel each other out. To see this, we note that the vector $w\equiv(U_I,U'_I)$ lies in $V^+$ so it is subject to the constraint \eqref{gentrans_horndeski}.  Since the only nonzero components of ${\cal P}_\text{GF}$ are ${\cal P}_\text{GF}(\xi)^{\mu\nu\rho\sigma}$, a calculation identical to \eqref{gf_cancel} establishes
\be
 {\cal P}_{\rm GF}(\xi)^{IJ}U_J = - (2 \xi_0 A_{\rm GF} + B_{\rm GF})^{IJ} T_J.
\ee
Contraction of the remaining terms in \eqref{2nd_row} with $T_I^*$ gives
\be
\label{TPU_inner}
T_I^*{\cal P}_\star(\xi)^{IJ} U_J=-T_I^*(2 \xi_0 A_\star + B_\star)^{IJ} T_{J}=(v,v)_+.
\ee
The symmetry of ${\cal P}_\star$ implies that the LHS can be written as $U {\cal P}_\star T^*=U ({\cal P}_\star T)^*$, which vanishes because ${\cal P}_\star T = {\cal P} T=0$ using the fact that $t_{\mu\nu}$ obeys \eqref{case1} and $T_I$ obeys \eqref{characteristic_horndeski}. Since the inner product on the RHS of \eqref{TPU_inner} is positive definite, it follows that $v=0$, which is a contradiction. Hence, $M(\xi_i)$ does not admit a non-trivial Jordan block in $V^+$, i.e., $M(\xi_i)$ is diagonalizable in $V^+$. The diagonalizability of $M(\xi_i)$ in $V^-$ follows similarly.

Since $M(\xi_i)$ has a basis of eigenvectors on the spaces $V^\pm$, it follows from the identity \eqref{Hsym2} that $H_\star$ is a symmetrizer on $V^\pm$. The definition of $H_\star$ shows that this symmetrizer is a smooth function of $\xi_i$ (even if the eigenvectors are not\footnote{As for a Lovelock theory, the physical eigenvectors may exhibit non-smoothness as a function of $\xi_i$ at values of $\xi_i$ for which two or more physical eigenvalues are degenerate.}). A symmetrizer on $V$ is now constructed from the (smooth) eigenvectors on $\tilde{V}^\pm$ and $\hat{V}^\pm$ and the (smooth) inner product $H_\star$ on $V^\pm$, just as for a weakly coupled Lovelock theory.

This concludes the proof of strong hyperbolicity for weakly coupled Horndeski theories.

\section{Discussion}
\label{sec:discuss}

\subsection{Application to numerical relativity}

Our modified harmonic gauge equations of motion involve two auxiliary Lorentzian inverse metrics $\tilde{g}^{\mu\nu}$ and $\hat{g}^{\mu\nu}$ as well as the physical inverse metric $g^{\mu\nu}$. The only conditions that we have imposed on these inverse metrics is that their causal cones should form a nested set as in Fig. \ref{fig:cones}. Our reasons for imposing these restrictions on the null cones are threefold: (i) we can ensure that our initial surface is spacelike w.r.t. all three metrics simultaneously; (ii) our proof of strong hyperbolicity of the gauge fixed equations requires that the null cones of the three metrics do not intersect; (iii) in GR our assumption that $g^{\mu\nu}$ has the innermost null cone (in the cotangent space), and hence $g_{\mu\nu}$ has the outermost null cone (in the tangent space), implies that the causal properties of the gauge fixed equations of motion are determined by the physical metric rather than either of the auxiliary metrics. 

Clearly there is considerable freedom in how we choose these metrics. A method that might be useful in numerical applications is as follows. Let $n_\mu$ be a unit (w.r.t. $g^{\mu\nu}$) normal to surfaces of constant $x^0$. We now choose
\be
\label{abdef}
 \tilde{g}^{\mu\nu} = g^{\mu\nu}- a n^\mu n^\nu \qquad \hat{g}^{\mu\nu} =  g^{\mu\nu} - b n^\mu n^\nu.
\ee
Our assumptions about the causal cones of the three metrics require that the functions $a(x)$ and $b(x)$ satisfy
\be\label{cond_ab}
0<a(x)<b(x) \qquad \qquad {\rm or} \qquad \qquad 0<b(x)<a(x).
\ee
This ensures that the null cones are nested either as in Fig. \ref{fig:cones} or as in Fig. \ref{fig:cones} with the null cones of $\tilde{g}^{\mu\nu}$ and $\hat{g}^{\mu\nu}$ interchanged. The simplest possibility would be to choose $a$ and $b$ to be constants satisfying the above inequalities. 

Although requirement (iii) is natural, it may not be essential for numerical relativity simulations. If one is willing to give up (iii) then the ordering of the causal cones in Fig. \ref{fig:cones} can be changed. 
If we interchange the null cones of $g^{\mu\nu}$ and $\tilde{g}^{\mu\nu}$ in Fig. \ref{fig:cones} then we have the alternative condition
\be
\label{cond_ab_2}
-1<a(x)<0<b(x)  
\ee
where the lower bound arises from the requirement that $\tilde{g}^{\mu\nu}$ is Lorentzian. In this case, causal properties of the gauge-fixed equation of motion will be determined by the unphysical metric $\tilde{g}^{\mu\nu}$, and we need to choose the initial lapse and shift to ensure that the initial surface is spacelike w.r.t. $\tilde{g}^{\mu\nu}$. 

For a Lovelock or Horndeski theory, strong hyperbolicity requires that the physical characteristics do not intersect the null cones of $\tilde{g}^{\mu\nu}$ and $\hat{g}^{\mu\nu}$, which will be the case at sufficiently weak coupling for any $a,b$ satisfying \eqref{cond_ab} or \eqref{cond_ab_2}. For stronger fields, one would not want a failure of strong hyperbolicity to arise simply from having chosen the null cones of $\tilde{g}^{\mu\nu}$ and $\hat{g}^{\mu\nu}$ too close to the null cone of ${g}^{\mu\nu}$, so $a,b$ should not be too close to zero. This should ensure that a failure of strong hyperbolicity (for stronger fields) arises from the behaviour of the physical degrees of freedom of the theory, rather than from the gauge fixing procedure. In a numerical simulation, one could check this by adjusting $a,b$ to see whether this extends the time for which the simulation runs.

Our formulation may have some advantages even for conventional GR. In numerical relativity, it might be possible to tailor the choice of $\tilde{g}^{\mu\nu}$ and $\hat{g}^{\mu\nu}$ to one's needs. For example, consider the choice \eqref{abdef} with $a = N/2-1$ where $N$ is the lapse function. Note that \eqref{cond_ab} requires $2<N<4$ whereas \eqref{cond_ab_2} allows $0<N<2$. In this case, the $\mu=0$ component of the modified harmonic gauge condition $H^\mu=0$ gives the so-called $1+\log$ slicing condition (where $N^k$ is the shift vector, $K$ is the trace of the extrinsic curvature $K_{ij}=-1/2~\mathcal{L}_n h_{ij}$)
 	\begin{equation} 
	(\partial_t-N^k\partial_k)N=-2KN
	\end{equation}
This is a popular choice of slicing due to its good singularity avoidance properties (see e.g. \cite{num_rel} and the references therein). 

As in conventional harmonic gauge, one has the option to introduce suitable source functions $F^\mu(x)$ and impose the generalized modified harmonic gauge condition 
\be
H^\mu\equiv -{\tilde g}^{\rho\sigma} \Gamma^\mu_{\rho\sigma}-F^\mu(x)=0.
\ee
Finally, the growth of numerical errors may be dealt with in the usual way \cite{Gundlach:2005eh}, i.e. by adding lower order homogeneous gauge fixing terms so that the equation of motion for the metric becomes
\be
	E^{\mu\nu}+{\hat P}_\alpha{}^{\beta\mu\nu}\partial_\beta H^\alpha
	-\kappa_1\left(n^\mu H^\nu+n^\nu H^\mu+\kappa_2~ n^\alpha H_\alpha~ g^{\mu\nu}\right)=0
	\ee
 where the constants $\kappa_1$, $\kappa_2$ are chosen so that constraint violations are damped away during the evolution.

\subsection{Domain of dependence}

In a Lovelock or Horndeski theory, the "physical" characteristic covectors (i.e. those associated with $V^\pm$) are generically non-null w.r.t. $g^{\mu\nu}$, with some of them spacelike and others timelike. (Some explicit examples have been studied in \cite{Reall2014}.) This means that causal properties of the theory are not determined by the null cone of $g_{\mu\nu}$. We will now discuss, in qualitative terms, the implications of this for the domain of dependence properties of these theories. We assume that the causal cones of the three metrics are related as in Fig. \ref{fig:cones}. 

Let $(M,g)$ be a spacetime satisfying the modified harmonic gauge Lovelock or Horndeski equation(s) of motion and the weakly coupled assumption. Let $\Sigma \subset M$ be an initial data surface, i.e., $\Sigma$ is spacelike w.r.t. $g^{\mu\nu}$ and is therefore non-characteristic (for sufficiently weak coupling, as explained above). Assume that the constraint equations and the harmonic gauge condition are satisfied on $\Sigma$. Then, as explained above, $g_{\mu\nu}$ will satisfy the original Lovelock/Horndeski equation(s) of motion in $\hat{D}(\Sigma) \subset M$. 

Now let $\Omega$ be a connected open subset of $\Sigma$. We define the domain of dependence of $\Omega$, denoted ${\cal D}(\Omega)$, to be the region of spacetime in which the solution does not change if we vary the initial data on $\Sigma\backslash \Omega$, keeping the data on $\Omega$ fixed. Strong hyperbolicity guarantees {\it local} well-posedness, which ensures that the solution is unique in a neighbourhood of $\Omega$ so ${\cal D}(\Omega)$ is non-empty. We define the Cauchy horizon of $\Omega$, denoted ${\cal H}(\Omega)$ to be the boundary of ${\cal D}(\Omega)$ in $M$. This will have two components: the future and past Cauchy horizons ${\cal H}^\pm(\Omega)$. We expect these to be the "innermost ingoing" characteristic hypersurfaces of \eqref{def_har_lovelock} emanating from, and tangential to, $\partial \Omega$, the boundary of $\Omega$.\footnote{See Ref. \cite{Reula:2004xd} for results supporting this expectation for the case of a quasilinear strongly hyperbolic system.} 

In modified harmonic gauge GR (or Einstein-scalar field theory), there are three types of characteristic surfaces, namely surfaces that are null w.r.t. one of the three inverse metrics. The ordering of the null cones assumed in Fig. \ref{fig:cones} implies that $D(\Omega) \subset \hat{D}(\Omega)$ and $D(\Omega) \subset \tilde{D}(\Omega)$, so it follows that the innermost ingoing characteristic hypersurface is null w.r.t. $g^{\mu\nu}$ and so ${\cal D}(\Omega) = D(\Omega)$, the domain of dependence defined w.r.t. the physical metric. So we have recovered the usual domain of dependence property of GR. 

In a weakly coupled modified harmonic gauge Lovelock/Horndeski theory, a characteristic surface is either null w.r.t. $\hat{g}^{\mu\nu}$ or w.r.t. $\tilde{g}^{\mu\nu}$ or its normal covector is associated with an eigenvector belonging to the "physical" space $V^\pm$. Generically, we expect $N=(1/2)d(d-3)$ distinct eigenvectors in each of $V^\pm$ and so generically there will be $N$ "physical" ingoing characteristic surfaces emanating from $\partial \Omega$. At weak coupling, the covectors normal to these surfaces will be timelike w.r.t. $\hat{g}^{\mu\nu}$ and $\tilde{g}^{\mu\nu}$ (since this is the case in GR), which implies that these physical characteristic surfaces are spacelike (i.e. "superluminal") w.r.t. $\hat{g}^{\mu\nu}$ and $\tilde{g}^{\mu\nu}$. Thus the Cauchy horizon of $\Omega$ will be one of these physical characteristic surfaces rather than one of the unphysical characteristic surfaces that is null w.r.t. $\hat{g}^{\mu\nu}$ or $\tilde{g}^{\mu\nu}$. 

We expect that, generically, this innermost ingoing physical characteristic surface will be spacelike also w.r.t. $g^{\mu\nu}$ (as for the examples in \cite{Reall2014}). Thus generically we expect the Cauchy horizon of $\Omega$ to be spacelike w.r.t. the physical metric $g^{\mu\nu}$. It would be interesting to study properties of this Cauchy horizon and the domain of dependence ${\cal D}(\Omega)$ in more detail.

\subsection*{Acknowledgments} 

We thank F. Abalos, P. Figueras and O. Reula for helpful conversations. ADK is supported by the George and Lilian Schiff Studentship. HSR is supported by STFC Grants PHY-1504541 and ST/P000681/1.

\begin{appendices}

\section{Hyperbolicity and well-posedness}\label{sec:hyperbolicity}

In this Appendix we review general results on the initial value problem of hyperbolic PDEs. First, we define the different notions of hyperbolicity appearing in the literature. Then, we state some technical theorems on how hyperbolicity is related to the well-posedness of the Cauchy problem. Our treatment of first order systems is based on \cite{Kreiss1989,Taylor91,Sarbach2012}.

\subsection{First order equations}

Let us consider a first order quasilinear equation on $\mathbb{R}^d$ of the form
		
		\begin{equation}\label{hyp_1}
		A(x,U)\partial_0 U+B^i(x,U)\partial_i U+C(x,U)=0
		\end{equation}

		\noindent
		in a coordinate system $x^\mu=(x^0,x^i)$ where $U$ is an $N$-component column vector, $A$, $B$ and $C$ are $N\times N$ matrix valued functions with smooth dependence on their arguments. We prescribe initial data $U(0,x^i)=f(x^i)$ on the hypersurface $x^0=0$. We assume that constant time surfaces are {\it non-characteristic}, that is, $A(x,U)$ is invertible on these slices. Then equation \eqref{hyp_1} can be rewritten as
		
		\begin{equation}\label{hyp_1v2}
		\partial_0 U=(B^{\prime})^i(x,U)\partial_i U+C^\prime(x,U)
		\end{equation}
		
		\noindent
		with $(B^{\prime})^i=-A^{-1}B^i$ and $C^\prime=-A^{-1}C$. Hyperbolicity is an algebraic condition on the matrix ${\cal M}(x,U,\xi_i)\equiv (B^{\prime})^i(x,U)\xi_i$ which contains information about the highest derivative (principal) terms in \eqref{hyp_1}.

		\begin{definition}
		{\it Let $\xi_i$ have unit norm w.r.t. a smooth positive definite (inverse) metric $G^{ij}$ on surfaces of constant $x^0$. Equation \eqref{hyp_1} is {\bf weakly hyperbolic} if all eigenvalues of ${\cal M}(x,U, \xi_i)$ are real for any such $\xi_i$. Equation \eqref{hyp_1} is {\bf strongly hyperbolic} if there exists an $N\times N$ Hermitian matrix valued function ${\cal K}(x,U,\xi_i)$ (called the {\bf symmetrizer}) that is positive definite with smooth dependence on its arguments, and a positive constant $\Lambda$ satisfying the conditions
				\begin{equation}\label{symm_eq}
		{\cal K}(x,U,\xi_i) {\cal M}(x,U,\xi_i)= {\cal M}^\dagger(x,U,\xi_i) {\cal K}(x,U,\xi_i)
		\end{equation}
		
		\noindent
		and
		
		\begin{equation}\label{bound_eq}
		\Lambda^{-1} I\leq {\cal K}(x,U,\xi_i) \leq \Lambda I.
		\end{equation}}
		\end{definition}

The standard way to prove strong hyperbolicity is to show that ${\cal M}(x,U,\xi_i)$ is diagonalizable with real eigenvalues and a complete set of linearly independent and bounded eigenvectors that depend smoothly on the variables $(x,U,\xi_i)$. Then one can use the eigenvectors to construct a suitable symmetrizer: if $S$ denotes the matrix whose columns are the eigenvectors of ${\cal M}(x,U,\xi_i)$, then it is easy to check that ${\cal K}=\left(S^{-1}\right)^\dagger S^{-1}$ is a positive definite, smooth and bounded symmetrizer.

Strong hyperbolicity can be similarly defined for first order equations that are not quasilinear in the spatial derivatives, i.e. equations of the form
		
	    \begin{equation}\label{nonlinear_1}
		\partial_0 U=B(x,U,\partial_i U).
		\end{equation}
		
		\noindent
		In this case, the condition of strong hyperbolicity refers to the matrix
		
		\be
		\label{cMdef}
		{\cal M}(x,U,\partial_i U,\xi_i)=(\partial_{\partial_j U}B)(x,U,\partial_i U)\xi_j
		\ee
		
		\noindent
		and the symmetrizer ${\cal K}(x,U,\partial_i U,\xi_i)$ must also depend smoothly on the spatial derivatives of $U$.
		
The following theorem relates strong hyperbolicity to the well-posedness of the initial value problem.
		
\begin{theorem}
For strongly hyperbolic first order quasilinear systems \eqref{hyp_1} and nonquasilinear systems of the form \eqref{nonlinear_1}, the Cauchy problem with initial data $U(0,x^i)=f(x^i)$ is well-posed in Sobolev spaces $H^s$ with $s>s_0$ for some constant $s_0$. That is to say, there exists a unique local solution $U\in C([0,T),\mathbb{R}^{d-1})$ with $T>0$ depending on the $H^s$-norm of the initial data.
\end{theorem}

The theorem above is proved in e.g. Chapter 5 of \cite{Taylor91}\footnote{Note in this reference strong hyperbolicity is called symmetrizable hyperbolicity.}. The statements about the quasilinear and the nonquasilinear systems differ only in the required order of regularity for the initial data, i.e. the value of $s_0$. However, we shall not be concerned with the problem of optimal regularity in this paper.

\subsection{Second order equations}

Let $x^\mu=(x^0,x^i)$ be coordinates in a $d$-dimensional spacetime 
Consider a set of fields $u_I$, $I = 1, \ldots, N$ satisfying a 
second order PDE 
\be
\label{2ndordera}
 V^I(x,u,\partial u, \partial^2 u ) = 0
\ee
with initial conditions
\be
\label{2nd_order_data2}
u(0,x^i)=f(x^i), \qquad \partial_0 u(0,x^i)=f^\prime(x^i)
\ee
We do not assume that the equation is quasilinear. We define the principal symbol as
\be 
 {\cal P}(\xi)^{IJ} \equiv {\cal P}^{IJ\mu\nu}\xi_\mu \xi_\nu \equiv \frac{\partial V^I}{\partial( \partial_\mu \partial_\nu u_J)} \xi_\mu \xi_\nu 
\ee 
Suppressing the $IJ$ indices we have
\be
{\cal P}(\xi)={\cal P}^{\mu\nu}\xi_\mu\xi_\nu=\xi_0^2 A+\xi_0 B(\xi_i)+C(\xi_i)
\ee
where the $N\times N$ matrices $A$, $B$, $C$ are given by
\be
 A={\cal P}^{00}, \qquad B(\xi_i)=2{\cal P}^{0i}\xi_i, \qquad C(\xi_i)={\cal P}^{ij}\xi_i\xi_j
\ee
These matrices depend on $x,u,\partial u$ and $\partial^2 u$ although we will not write this explicitly. {\it We will restrict attention to equations for which $V^I$ depends linearly on $\partial_0^2 u$.} This implies that 
\be
\label{Wdef}
 V^I(x,u,\partial u, \partial^2 u ) = A^{IJ}(x,u,\partial u, \partial_0 \partial_i u, \partial_i \partial_j u) \partial_0^2 u_J + W^I (x,u,\partial u, \partial_0 \partial_i u, \partial_i \partial_j u)
\ee
We also assume that $x^0=0$ is non-characteristic, which means that initial data \eqref{2nd_order_data2} is chosen so that $\det A \ne 0$ at $x^0=0$. By continuity, this condition will continue to hold in a neighbourhood of $x^0=0$. We can then rewrite the equation as
\be
\label{2ndordereofm}
 \partial_0^2 u_I = X_I( x,u,\partial u, \partial_0 \partial_i u, \partial_i \partial_j u)
\ee
where
\be
 X_I = - (A^{-1})_{IJ} W^J
\ee 
Note that
\bea
\label{dX1}
 \frac{\partial X_I}{\partial ( \partial_i \partial_j u_J)} &=& - (A^{-1})_{IK} \frac{\partial W^K}{\partial ( \partial_i \partial_j u_J)} + (A^{-1})_{IK} \frac{\partial A^{KL}}{\partial ( \partial_i \partial_j u_J)} (A^{-1})_{LM} W^M \nonumber \\
 &=& -(A^{-1})_{IK} \left[  \frac{\partial W^K}{\partial ( \partial_i \partial_j u_J)} + \frac{\partial A^{KL}}{\partial ( \partial_i \partial_j u_J)} \partial_0^2 u_L \right] = -(A^{-1})_{IK} \frac{\partial V^K}{\partial ( \partial_i \partial_j u_J)} \nonumber \\
 &=& - (A^{-1})_{IK} {\cal P}^{KJij}
\eea
where we used the equation of motion \eqref{2ndordereofm} in the second equality. Similarly\footnote{The factor of $2$ can be understood by varying $\partial_\mu \partial_\nu u$ in \eqref{Wdef}. On the LHS this gives ${\cal P}^{\mu \nu} \delta( \partial_\mu \partial_\nu u) $ which, by the symmetry of mixed partial derivatives, contains a term $2 {\cal P}^{0i}\delta( \partial_0 \partial_i u)$.}
\be
\label{dX2}
 \frac{\partial X_I}{\partial ( \partial_0 \partial_i u_J)} = -2(A^{-1})_{IK} {\cal P}^{KJ0i}.
\ee

We will now write the second order system \eqref{2ndordereofm} as a first order system. Define $v_0\equiv \partial_0 u$ and $v_i\equiv \partial_i u$. Then \eqref{2ndordereofm} implies 
\bea\label{1stordereofm}
\partial_0 u&=&v_0 \nonumber\\
\partial_0 v_i&=&\partial_i v_0 \nonumber \\
\partial_0 v_0&=&-X(x,u,v,\partial_i v_0,\partial_i v_j)
\eea
together with the constraint equations ${\cal C}_i=0$ where
\be
\label{aux_constr2}
{\cal C}_i\equiv v_i-\partial_i u.
\ee
The evolution system \eqref{1stordereofm} is of the form \eqref{nonlinear_1} with the variable $U\equiv (u,v_i,v_0)$. Initial data for the auxiliary variable $v_i$ can be constructed from the data \eqref{2nd_order_data2} by taking the spatial derivatives of the data for $u$ on the initial surface. Hence the initial data for $U$ is 
\be
\label{1storderdata}
U(0,x^j)=\left(f(x^j), \partial_i f(x^j),f^\prime(x^j)\right).
\ee
Note in particular that by construction this data satisfies the constraint equation. Then it follows 
from the equations \eqref{1stordereofm} that the constraints will also be satisfied at later times. To see this take a time derivative of \eqref{aux_constr2} which gives
\be\label{aux_constr_prop2}
\partial_0 {\cal C}_i=\partial_0 v_i-\partial_0\partial_i u=\partial_0 v_i-\partial_i v_0=0.
\ee
It follows that if $(u,v_0,v_i)$ is a solution of \eqref{1stordereofm} arising from initial data of the form \eqref{1storderdata} then $u$ will be a solution of \eqref{2ndordereofm} satisfying the initial conditions \eqref{2nd_order_data2}.

We will now demand that the first order system \eqref{1stordereofm} is strongly hyperbolic and determine the conditions that this imposes on the second order system that we started from. The matrix ${\cal M}$ defined in \eqref{cMdef} is
\be
{\cal M} = \left(\begin{array}{ccc} 0 & 0 & 0 \\ 0  & 0 & \xi_i  \\ 0 & \frac{\partial X_I}{\partial (\partial_i (v_j)_J)}\xi_i& \frac{\partial X_I}{\partial (\partial_i (v_0)_J)}\xi_i   \end{array} \right)= \left(\begin{array}{ccc} 0 & 0 & 0 \\ 0  & 0 & \xi_i  \\ 0 & -(A^{-1})_{IK} {\cal P}^{KJij} \xi_i& - (A^{-1})_{IK} B^{KJ} \end{array} \right)
\ee
in the second equality we used $\partial X/(\partial (\partial_i v_j)) = \partial X/(\partial (\partial_i \partial_j u ))$ and equation \eqref{dX1}, $\partial X/(\partial (\partial_i v_0)) = \partial X/(\partial (\partial_0 \partial_i u ))$ and equation \eqref{dX2}, and the definition of $B^{IJ}$. 

This matrix acts on a vector space of dimension $(d+1)N$. A general vector in this space can be written $(s_J,(t_i)_J,(t_0)_J)$. The definition of strong hyperbolicity of the first order system refers to a smooth Riemannian (inverse) metric $G^{ij}$ on surfaces of constant $x^0$. It is convenient to separate $(t_i)_J$ into a part $(t^\perp_i)_J$ perpendicular to $\xi_i$ w.r.t. $G^{ij}$, and a part $t^\parallel_J \xi_i$ parallel to $\xi_i$. We then order the components of our vector as $(s_J,(t^\perp_i)_J,t^\parallel_J,(t_0)_J)$. With this decomposition, ${\cal M}$ becomes
\be\label{cal_M_decomp}
{\cal M} = \left(\begin{array}{cccc} 0 & 0 & 0 & 0 \\ 0  & 0 & 0 & 0 \\ 0 & 0 & 0 & I  \\ 0 & -(A^{-1})_{IK} ({\cal P}^{KJij} \xi_i)^\perp& -(A^{-1})_{IK} C^{KJ} & -(A^{-1})_{IK} B^{KJ} \end{array} \right)
\ee
where $I$ is the $N \times N$ identity matrix and $({\cal P}^{KJij} \xi_i)^\perp$ is the restriction of ${\cal P}^{KJij} \xi_i$ to vectors of the form. $(t^\perp_i)_J$.

We write a general vector as the sum of $(s_J,(t^\perp_i)_J,0,0)$ and $(0,0,t^\parallel_J,(t_0)_J)$ respectively. This gives a block decomposition of ${\cal M}$
\be
{\cal M} = \left(\begin{array}{cc}  0 & 0 \\ L & M \end{array} \right)
\ee
where $L$ is the $2N \times (d-1) N$ matrix
\be
 L = \left( \begin{array}{cc} 0 & 0 \\ 0 & -(A^{-1})_{IK} ({\cal P}^{KJij} \xi_i)^\perp \end{array} \right) 
\ee
and $M$ is the $2N \times 2N$ matrix
\be 
\label{Mdefinition}
 M  = \left(\begin{array}{cc} 0 & I \\ - A^{-1}C & -A^{-1} B \end{array} \right)
\ee
The matrices $L$ and $M$ (and hence ${\cal M}$) depend on $(x,u,\partial u, \partial_0 \partial_i u, \partial_i \partial_j u,\xi_i)$, or equivalently on $(x,u,v,\partial_i v,\xi_i)$.

We now examine the conditions for our first order system to admit a symmetrizer. Any Hermitian $(d+1)N \times (d+1)N$ matrix ${\cal K}$ can be written as
\be\label{cal_K}
 {\cal K} = \left( \begin{array}{cc} E & F \\ F^\dagger & K \end{array} \right)
\ee
where $E$ is a $(d-1)N \times (d-1)N$ Hermitian matrix, $K$ is a $2N \times 2N$ Hermitian matrix and $F$ is a $(d-1)N \times 2N$ matrix.  Equation \eqref{symm_eq} reduces to
\be
\label{KMeq}
 KM = M^\dagger K
\ee
and
\be
\label{Feqs}
 FL = (FL)^\dagger \qquad \qquad FM  =L^\dagger K
\ee
Let us assume that we can find a positive definite Hermitian matrix $K$, depending smoothly on $(x,u,v,\partial_i v,\xi_i)$, and satisfying \eqref{KMeq} and $\lambda^{-1} I \le K \le \lambda I$ for some positive constant $\lambda$. Assume also that $M$ is invertible. We can then satisfy \eqref{Feqs} with $F = L^\dagger K M^{-1}$. We then take $E = c I$ where $c$ is a positive constant. Let $T\equiv(T_1,T_2)^\dagger$ with $T_1\equiv(s_J,(t^\perp_i)_J)^\dagger$ and $T_2\equiv (t^\parallel_J,(t_0)_J)^\dagger$ and consider
\be
T^\dagger {\cal K} T=c T_1^\dagger T_1+T_1^\dagger F T_2+T_2^\dagger F^\dagger T_1+T_2^\dagger K T_2.
\ee
If ${\cal P}^{IJ\mu\nu}$ is uniformly bounded then so are the matrices $M$, $L$ and $F$ when $\xi_i$ is a unit covector w.r.t. a positive definite (inverse) metric $G^{ij}$. By taking $c$ large enough we can ensure that ${\cal K}$ is positive definite and \eqref{bound_eq} holds for some $\Lambda$. Hence we have constructed a symmetrizer for the first order system. 

This motivates the following definitions for our second order equation:
\begin{definition}
{\it Consider a second order PDE \eqref{2ndordera} satisfying \eqref{Wdef}. Define the matrix $M$, depending on $(x,u,\partial u, \partial_0 \partial_i u, \partial_i \partial_j u, \xi_i)$, by \eqref{Mdefinition}. Let $\xi_i$ have unit norm w.r.t. a smooth positive definite (inverse) metric $G^{ij}$ on surfaces of constant $x^0$. The equation is {\bf weakly hyperbolic} if all eigenvalues of $M$ are real for any such $\xi_i$. The equation is {\bf strongly hyperbolic} if there exists a $2N\times 2N$ Hermitian matrix valued function $K(x,u,\partial u, \partial_0 \partial_i u, \partial_i \partial_j u, \xi_i)$ (called the {\bf symmetrizer}) that is positive definite with smooth dependence on its arguments, and that satisfies \eqref{KMeq}, and a positive constant $\lambda$ such that $\lambda^{-1} I \le K \le \lambda I$. 
}	

Note that any eigenvalue of ${\cal M}$ is either $0$ or an eigenvalue of $M$. Hence if our second order PDE is weakly hyperbolic according to the above definition then the first order system \eqref{1stordereofm} is weakly hyperbolic according to our previous definition. If our second order PDE is strongly hyperbolic according to the above definition then and $M$ is also invertible then the above discussion shows that we can construct a symmetrizer for the first order system \eqref{1stordereofm} and so this system is also strongly hyperbolic. Hence we can apply the well-posedness theorem stated at the end of the previous section to deduce well-posedness of the initial value problem for the second order system. 

The condition that $M$ should be invertible was used above but it is not a necessary condition for well-posedness. However, if this condition is not satisfied then it may be necessary to explore different ways of reducing the second order equation to a first order system. Therefore we will assume that this condition is satisfied. This condition is equivalent to the condition that $C$ should be invertible, which is equivalent to the condition that $(0,\xi_i)$ is never characteristic, which is equivalent to the condition that $\xi_0 \ne 0$ for any characteristic covector $\xi_\mu$. 

In our modified harmonic gauge formulation of $GR$ we choose our lapse and shift so that $\partial/\partial x^0$ is timelike w.r.t. the metric with the innermost causal cone and hence timelike w.r.t. all three metrics. This implies that a covector with $\xi_0=0$ is spacelike w.r.t. all three (inverse) metrics and hence non-characteristic so $M$ is non-degenerate. By continuity, this condition will also be satisfied by weakly coupled Lovelock and Horndeski theories. 

\end{definition}

\section{Maxwell theory in modified Lorenz gauge}

\label{app:maxwell}

In this section we will discuss the analogue of our new formulation of GR in the simpler setting of Maxwell theory. The presentation follows closely our discussion of modified harmonic gauge GR in section \ref{sec:GR}.

\subsection{Equation of motion}

The vacuum Maxwell equations for the (antisymmetric) electromagnetic field tensor $F^{\mu\nu}$ on a globally hyperbolic ($d$ dimensional) spacetime $(M,g)$ are
\be\label{maxwell_eom}
E^\mu=0
\ee
with
\be
E^\mu\equiv \nabla_\nu F^{\mu\nu}
\ee
and the Bianchi identity for $F$
\be
\nabla_{[\rho}F_{\mu\nu]}=0
\ee
which implies that $F_{\mu\nu}$ can be written (locally) as the exterior derivative of a potential $1$-form $A$:
\be
F_{\mu\nu}=\nabla_\mu A_\nu-\nabla_\nu A_\mu.
\ee
Let ${\tilde g}^{\mu\nu}$ be an (inverse) Lorentzian metric on $M$ and let
\be
H\equiv -{\tilde g}^{\alpha\beta}\nabla_\beta A_\alpha
\ee
Our modified Lorenz gauge condition is then
\be\label{mod_lorenz_cond}
H=0.
\ee
We introduce another (inverse) Lorentzian metric ${\hat g}^{\mu\nu}$ to write the gauge-fixed Maxwell equations as
\be\label{gf_eom_maxwell}
{E}^\mu_{\rm mLg}=0
\ee
where ${E}^\mu_{\rm mLg}$ is defined by
\be
{E}^\mu_{\rm mLg}=E^\mu+{\hat g}^{\mu\nu}\nabla_\nu H.
\ee
Now the gauge-fixed equations for $A_\mu$ can be written as 
\be
{E}^\mu_{\rm mLg}\equiv -g^{\mu\nu}g^{\rho\sigma}\nabla_\rho\nabla_\sigma A_\nu+g^{\mu\rho}\nabla_\rho \left(g^{\nu\sigma}\nabla_\sigma A_\nu\right)-{\hat g}^{\mu\rho}\nabla_\rho \left({\tilde g}^{\nu\sigma}\nabla_\sigma A_\nu\right)=0
\ee
In conventional Lorenz gauge $g^{\mu\nu}={\hat g}^{\mu\nu}={\tilde g}^{\mu\nu}$ and the second and third term in the above equation cancel, leaving a simple wave equation for $A$.

We assume that the null cones of the three metrics are related as in Fig. \ref{fig:cones}. 

\subsection{Propagation of the gauge condition}

To show that the gauge condition \eqref{mod_lorenz_cond} is propagated we follow a standard argument similar to the original Lorenz gauge case (and analogous to the harmonic gauge GR case). We assume that we have a solution of \eqref{gf_eom_maxwell} on $M$ and consider the identity 
\be\label{maxwell_bianchi}
\nabla_\mu E^\mu=\nabla_\mu\nabla_\nu F^{\mu\nu}=0
\ee
that is a consequence of the antisymmetry of $F^{\mu\nu}$. Using this we have
\be
\label{Heqmax}
0=\nabla_\mu{E}^\mu_{\rm mLg}={\hat g}^{\mu\nu}\nabla_\mu\nabla_\nu H+\left(\nabla_\mu {\hat g}^{\mu\nu}\right)\nabla_\nu H
\ee
which is a homogeneous linear wave equation for $H$.

Given initial data specified on a surface $\Sigma$ spacelike w.r.t. $g^{\mu\nu}$ and hence also w.r.t. $\hat{g}^{\mu\nu}$ and $\tilde{g}^{\mu\nu}$, one can impose the gauge condition $H=0$ initially and then the constraint equation $n_\mu E^\mu=0$ (where $n_\mu$ is normal to the hypersurface) implies that $n\cdot \partial H=0$ initially. Hence, by well-posedness of \eqref{Heqmax}, $H$ vanishes throughout $\hat{D}(\Sigma)$ and therefore also throughout $D(\Sigma) \subset \hat{D}(\Sigma)$. So any solution of \eqref{gf_eom_maxwell} arising from initial data satisfying the constraint equation and the gauge condition will also satisfy $E^\mu=0$ in $D(\Sigma)$.

\subsection{Strong hyperbolicity}

The principal symbol of \eqref{gf_eom_maxwell} acting on a covector $t_\mu$ is given by
\be
{\cal P}(\xi)^{\mu\nu}t_\nu={\cal P}_\star(\xi)^{\mu\nu}t_\nu+{\cal P}_\text{GF}(\xi)^{\mu\nu}t_\nu
\ee
with
\be
{\cal P}_\star(\xi)^{\mu\nu}t_\nu=-g^{\mu\nu}g^{\gamma\delta}\xi_\gamma\xi_\delta t_\nu+g^{\mu\gamma}g^{\nu\delta}\xi_\gamma\xi_\delta t_\nu
\ee
and
\be
{\cal P}_\text{GF}(\xi)^{\mu\nu}t_\nu=-{\hat g}^{\mu\gamma}{\tilde g}^{\nu\delta}\xi_\gamma\xi_\delta t_\nu
\ee
As for GR, our task is to show that $M(\xi_i)$ has real eigenvalues and possesses a complete set of eigenvectors with smooth dependence on the unit vector $\xi_i$. This boils down to studying the characteristic equation
\be
P(\xi)^{\mu\nu}t_\nu=0.
\ee
In analogy with the GR case, the identity \eqref{maxwell_bianchi} implies that for any $\xi_\mu$
\be
\xi_\mu{\cal P}_\star(\xi)^{\mu\nu}=0.
\ee
Therefore, by considering
\be
0=\xi_\mu{\cal P}(\xi)^{\mu\nu}t_\nu=\xi_\mu{\cal P}_\text{GF}(\xi)^{\mu\nu}t_\nu=\left({\hat g}^{\mu\gamma}\xi_\mu\xi_\gamma\right)\left({\tilde g}^{\nu\delta}\xi_\delta t_\nu\right)
\ee
we find that at least one of the following two cases must hold: (i) ${\tilde g}^{\nu\delta}\xi_\delta t_\nu=0$ or (ii) ${\hat g}^{\mu\gamma}\xi_\mu\xi_\gamma=0$.

The physical interpretation of case (i) is a high-frequency wave with polarization $t_\mu$ and wave vector $\xi_\mu$ that satisfies the modified Lorenz gauge condition. Case (i) can be split into two subcases: characteristics falling into this category must have either (ia) $g^{\mu\nu}\xi_\mu\xi_\nu\neq 0$ or (ib) $g^{\mu\nu}\xi_\mu\xi_\nu= 0$. An argument very similar to the one presented in the GR case reveals that characteristics in subcase (ia) satisfy ${\tilde g}^{\mu\nu}\xi_\mu\xi_\nu=0$ which has two real solutions for $\xi_0$ depending smoothly on $\xi_i$. Adopting the notation used in section \ref{sec:GR_strong_hyperbolicity}, we shall denote these solutions by ${\tilde \xi}_0^\pm$. The corresponding polarizations are $t_\mu^\pm=\xi_\mu^\pm\equiv ({\tilde \xi}_0^\pm,\xi_i)$; these are "pure gauge" vectors associated with the residual gauge freedom in \eqref{mod_lorenz_cond}. On the other hand, the modes of subcase (ib) have eigenvalues $\xi_0^\pm$ (obtained by solving $g^{\mu\nu}\xi_\mu\xi_\nu= 0$ for $\xi_0$). The only requirement on the corresponding eigenvectors is that they be orthogonal to $\xi_\mu^\pm=(\xi_0^\pm,\xi_i)$ in both ${\tilde g}^{\mu\nu}$ and $g^{\mu\nu}$. Hence there are $d-2$ linearly independent eigenvectors in this subclass (for each sign choice in $\xi_0^\pm$). These can be interpreted as the physical photon polarizations. Finally, case (ii) contains the "gauge condition violating" modes whose characteristic covectors denoted by ${\hat\xi}_\mu^\pm=({\hat \xi}_\mu^\pm,\xi_i)$ are null w.r.t. ${\hat g}^{\mu\nu}$. For each of these covectors there is only one corresponding eigenvector that can be found by following a similar strategy as in the GR case. In summary, we find that $M(\xi_i)$ has $2d$ real eigenvalues and a complete set of smooth and real eigenvectors which guarantees that \eqref{gf_eom_maxwell} is strongly hyperbolic.

Strong hyperbolicity of our modified harmonic gauge formulation of GR is robust against deformation of the theory into a weakly coupled Lovelock or Horndeski theory. Similarly, strong hyperbolicity of the above formulation of Maxwell theory should be robust against the inclusion of small nonlinear terms, i.e., a weakly coupled theory of nonlinear electromagnetism (with second order equations of motion). The initial value problem for such theories has been studied by writing the equations of motion as a first order system for the field strength tensor $F_{\mu\nu}$. This system is {\it symmetric} hyperbolic under certain conditions \cite{nonlinear_ed}. Our modified Lorenz gauge would provide an alternative well-posed formulation of such theories, with equations written in terms of the potential $A_\mu$ instead of the field strength.

\end{appendices}

%\bibliographystyle{JHEP}
%\bibliography{Remote}

\begin{thebibliography}{0}


\bibitem{Lovelock1971}
	D. Lovelock, {\it The Einstein Tensor and Its Generalizations}, \href{http://scitation.aip.org/content/aip/journal/jmp/12/3/10.1063/1.1665613}{J. Math. Phys. {\bf 12} 498 (1971)}

\bibitem{Burgess:2003jk}
     C. P. Burgess, {\it Quantum gravity in everyday life: General relativity as an effective field theory}, Living Rev. Rel. {\bf 7} (2004) 5-56, arXiv:gr-qc/0311082

\bibitem{Horndeski1974}
	G. W. Horndeski, {\it Second-order scalar-tensor field equations in a four-dimensional space}, \href{http://link.springer.com/article/10.1007/BF01807638}{Int. J. Theor. Phys. {\bf 10} (1974) 363-384}

\bibitem{Weinberg:2008hq}
S. Weinberg, {\it Effective Field Theory for Inflation}, Phys. Rev. {\bf D77} (2008) 123541, arXiv:0804.4291 [hep-th].

\bibitem{Kobayashi:2011nu}
	T. Kobayashi, M. Yamaguchi, and J. Yokoyama, {\it Generalized G-Inflation: Inflation with the Most General Second-Order Field Equations}, \href{http://ptp.oxfordjournals.org/content/126/3/511.abstract}{Prog. Theor. Phys., {\bf 126} (2011) 511-529}, arXiv:1105.5723 [gr-qc].

\bibitem{Kreiss1989}
	H.-O. Kreiss and J. Lorenz, {\it Initial-boundary value problems and the Navier-Stokes equations}, \href{http://epubs.siam.org/doi/abs/10.1137/1.9780898719130.appc}{{\bf 136} (1989).}


\bibitem{Taylor91}
	M. E. Taylor, {\it Pseudodifferential Operators and Nonlinear PDE}, (Birkh{\"{a}}user, Boston, MA, 1991).


\bibitem{Sarbach2012}
	O. Sarbach and M. Tiglio, {\it Continuum and Discrete Initial-Boundary Value Problems and Einstein's Field Equations}, Living Rev. Relativ. {\bf 15} (2012) 9, \href{http://arxiv.org/abs/1203.6443}{arXiv:1203.6443 [gr-qc].}

\bibitem{Reall2014}
	H. S. Reall, N. Tanahashi and B. Way, {\it Causality and hyperbolicity of Lovelock theories}, \href{http://stacks.iop.org/0264-9381/31/i=20/a=205005?key=crossref.8603cb2d17fc2fe84d3c2b9abe452d10}{Class. Quantum Gravity {\bf 31} (2014) 205005}, arXiv:1406.3379 [gr-qc]


\bibitem{Papallo:2017qvl}
      G. Papallo and H. S. Reall, {\it On the local well-posedness of Lovelock and Horndeski theories}, Phys. Rev. {\bf D96} (2017) 044019, arXiv:1705.04370 [gr-qc].

\bibitem{Ripley:2019hxt}
     J. L. Ripley and F. Pretorius, {\it Hyperbolicity in Spherical Gravitational Collapse in a Horndeski Theory}, Phys. Rev. {\bf D99} (2019), no. 8. 084014, arXiv:1902.01468 [gr-qc].

\bibitem{Ripley:2019irj}
      J. L. Ripley and F. Pretorius, {\it Gravitational collapse in Einstein-dilaton-Gauss-Bonnet gravity}, Class. Quant. Grav. {\bf 36} (2019), no. 13. 134001, arXiv:1903.07543 [gr-qc].

\bibitem{Ripley:2019aqj}
      J. L. Ripley and F. Pretorius, {\it Scalarized Black Hole dynamics in Einstein dilaton Gauss-Bonnet Gravity}, Phys. Rev. {\bf D101} (2020) no. 4. 044015, arXiv:1911.11027 [gr-qc].

\bibitem{nonlinear_ed}
  {F. Abalos, F. Carrasco, E. Goulart and O. Reula}, {\it Nonlinear electrodynamics as a symmetric hyperbolic system}, Phys. Rev. {\bf D92} (2015), no. 8. 084024.

\bibitem{Wald:1984rg}
      R. M. Wald, {\it General Relativity}, Chicago Univ. Pr., Chicago, USA, 1984.


\bibitem{Papallo:2017ddx}
      G. Papallo, {\it On the hyperbolicity of the most general Horndeski theory}, Phys. Rev. {\bf D96} (2017) no. 12. 124036, arXiv:1710.10155 [gr-qc].


\bibitem{Kovacs:2019jqj}
      A. D. Kov\'acs, {\it Well-posedness of cubic Horndeski theories}, Phys. Rev. {\bf D100} (2019) no 2. 024005, arXiv:1904.00963 [gr-qc].


\bibitem{Allwright:2018rut}
      G. Allwright and L. Lehner, {\it Towards the nonlinear regime in extensions to GR: assessing possible options}, Class. Quant. Grav. {\bf 36} (2019), no. 8, 084001, arXiv:1808.07897 [gr-qc].


\bibitem{Witek:2018dmd}
      H. Witek, L. Gualtieri, P. Pani and T. P. Sotiriou, {\it Black holes and binary mergers in scalar Gauss-Bonnet gravity: scalar field dynamics}, Phys. Rev. {\bf D99} (2019), no. 6, 064035, arXiv:1810.05177 [gr-qc].

\bibitem{Okounkova:2019zep}
      M. Okounkova, {\it Stability of Rotating Black Holes in Einstein Dilaton Gauss-Bonnet Gravity}, Phys. Rev. {\bf D100}, (2019) no. 12. 124054, arXiv:1909.12251 [gr-qc].


\bibitem{Okounkova:2020rqw}
      M. Okounkova, {\it Numerical relativity simulation of GW150914 in Einstein dilaton Gauss-Bonnet gravity}, (2020) arXiv:2001.03571 [gr-qc].

\bibitem{Okounkova:2017yby}
M. Okounkova, L. C. Stein, M. A. Scheel, and D. A. Hemberger, {\it Numerical binary black hole mergers in dynamical Chern-Simons gravity: Scalar field}, Phys. Rev. {\bf D96} (2017), 044020, [arXiv:1705.07924].

\bibitem{Flanagan:1996gw}
     E. E. Flanagan and R. M. Wald, {\it Does back reaction enforce the averaged null energy condition in semiclassical gravity?}, Phys. Rev. {\bf D54} (1996) 6233-6283, arXiv:gr-qc/9602052.

\bibitem{letter}
      A. D. Kovacs and H. S. Reall, {\it Well-posed formulation of scalar-tensor effective field theory}, (2020) arXiv:2003.04327 [gr-qc].

\bibitem{Choquet-Bruhat1988}
	Y. Choquet-Bruhat, {\it The Cauchy problem for stringy gravity}, \href{http://scitation.aip.org/content/aip/journal/jmp/29/8/10.1063/1.527841}{J. Math. Phys. {\bf 29}} (1988) 1891.


\bibitem{Aragone:1987jm}
      C. Aragone, {\it Stringy characteristics of effective gravity}, in SILARG 6: 6th Latin American Symposium on Relativity and Gravitation Rio de Janeiro, Brazil, July 13-18, 1987, 60-69, (1987).


\bibitem{Kato1976}
	T. Kato, {\it Perturbation Theory for Linear Operators}, Classics in Mathematics, Springer Berlin Heidelberg, 1976.


\bibitem{num_rel}
 E. Gourgoulhon, {\it 3+1 Formalism in General Relativity}, vol. 846. Springer, Berlin, Heidelberg, 2012.

\bibitem{Gundlach:2005eh}
C. Gundlach, J. M. Martin-Garcia, G. Calabrese, and I. Hinder, {\it Constraint damping in the Z4 formulation and harmonic gauge}, Class. Quant. Grav. {\bf 22} (2005) 3767-3774, [gr-qc/0504114].

\bibitem{Reula:2004xd}
O. A. Reula, {\it Strongly hyperbolic systems in general relativity}, Diff. Eq.{\bf 01} (2004) 251, [gr-qc/0403007].

\end{thebibliography}

\end{document}